\documentclass[12pt]{article}

\pdfoutput=1

\usepackage{graphics}
\usepackage{amssymb}
\usepackage{amsmath}
\usepackage{hyperref}
\usepackage{bbold}
\usepackage{tikz}

\newcommand{\beq}{\begin{equation}}

\newcommand{\eeq}{\end{equation}}
\newcommand{\bea}{\begin{eqnarray}}
\newcommand{\eea}{\end{eqnarray}}

\begin{document}

\begin{center}
${}$\\
\vspace{100pt}
{ \Large \bf Quantum Curvature as
\\ \vspace{10pt} Key to the Quantum Universe 
}

\vspace{36pt}

{\sl R. Loll}

\vspace{18pt}
{\footnotesize

Institute for Mathematics, Astrophysics and Particle Physics, Radboud University \\ 
Heyendaalseweg 135, 6525 AJ Nijmegen, The Netherlands.\\ 

\vspace{5pt}
{\it and}\\
\vspace{5pt}

Perimeter Institute for Theoretical Physics,\\
31 Caroline St N, Waterloo, Ontario N2L 2Y5, Canada.\\
}
\vspace{24pt}

\end{center}


\begin{center}
{\bf Abstract}
\end{center}

\noindent 
Curvature is a key notion in General Relativity, characterizing the local physical properties of spacetime. 
By contrast, the concept of curvature has received scant attention in nonperturbative quantum gravity.
One may even wonder whether in a Planckian regime meaningful notions of (quantum) curvature exist at all.
Remarkably, recent work in quantum gravity using Causal Dynamical Tri\-an\-gulations (CDT) has demonstrated both
the existence and usefulness of a new notion of {\it quantum Ricci curvature (QRC)}, which relies neither on smooth structures nor on tensor calculus. 
This overview article recalls some classical notions related to curvature and parallel transport, as well as previous unsuccessful attempts to construct 
quantum curvature observables based on deficit angles and Wilson loops. It introduces the quasi-local QRC on
piecewise flat triangulations, and describes its behaviour in a purely classical setting, its use in quantum observables,
and currently known results in (C)DT quantum gravity in two and four dimensions. 
The QRC opens the door to a range of interesting physical observables that were previously out of reach, and will help to 
bridge the gap between the nonperturbative quantum theory and gravitational phenomena at lower energies.

\vspace{12pt}
\noindent

\newpage

\section{Introduction}
\label{sec:intro}

Anyone who starts learning about the mathematical structure of General Relativity quickly understands that the notion of {\it curvature}
plays a pivotal role in the theory. Curvature, which describes how a spacetime deviates locally from being flat, appears on the left-hand side of the 
Einstein equations 
\begin{equation}
R_{\mu\nu}(x)-\frac{1}{2}g_{\mu\nu}(x)R(x)=8 \pi G_N T_{\mu\nu}(x)
\label{ein}
\end{equation}
in the form of the Ricci tensor $R_{\mu\nu}(x)$ and the Ricci scalar $R(x)$, both obtained by contractions of the full Riemann
curvature tensor $R^\kappa{}_{\lambda\mu\nu}(x)$. The Einstein equations relate the curvature
to the distribution of matter and energy, as captured by the energy-momentum tensor $T_{\mu\nu}$ on the right-hand 
side of eq.\ (\ref{ein}). The fact that we model curved spacetimes in General Relativity by differentiable manifolds $M$
equipped with a metric tensor $g_{\mu\nu}$ goes back to the foundational work of Riemann in the mid-19th century. 
As described in his famous habilitation thesis \cite{Riemann1854}, he was guided by ``experience" -- i.e.\ his physical, Newtonian
intuition -- to introduce the infinitesimal line element
\begin{equation}
ds^2=g_{\mu\nu}(x)dx^\mu dx^\nu.
\label{line}
\end{equation}
The immense power of the formalism of Riemannian geometry, when generalized to four-dimensional Lorentzian (pseudo-Riemannian) spacetimes,
is reflected in the spectacular successes of General Relativity, which has given us black holes, dynamical universes,
gravitational waves and more.   
Remarkably from a modern point of view, Riemann already anticipated that ``[his] considerations may not apply in the immeasurably small", and
may have to be revisited in the light of new physical observations. 

This is precisely the situation we find ourselves in when trying to
understand and describe the microscopic structure of spacetime in a nonperturbative quantum theory of gravity. 
Although we cannot currently observe
physics at the Planck scale directly, our understanding of the nature of quantum fields at very short distances, acquired since Riemann's days, 
leads us to believe that the metric fields of the classical formulation may not be adequate to describe the properties of quantum spacetime
in the vicinity of the Planck scale. Generalizing or altogether abandoning the metric (or an equivalent field, like the vierbein (tetrad) $e^\mu_a(x)$
of a first-order formulation of gravity) in these circumstances raises the question of what happens to curvature.  

In this work, we will examine the quest for curvature in a relatively conservative quantum scenario, where a metric structure is still present 
but is not smooth. The concrete setting we will be working with is lattice gravity in terms of Causal Dynamical Triangulations (CDT), where 
the gravitational path integral is defined nonperturbatively as a continuum limit of a sum over piecewise flat triangulated spaces 
\cite{review1,review2}. However, given the simplicity of the ingredients,
many of our considerations and constructions are likely applicable in other regularized or discretized formulations of quantum gravity too. 
The centrepiece of our exposition is the {\it quantum Ricci curvature},
a recently introduced notion of curvature \cite{Klitgaard2017,Klitgaard2018}, whose novelty is twofold: 
it has been used to construct quantum curvature observables that 
have been shown to be well defined and finite in the quantum theory, and it is a quantum implementation of genuine Ricci curvature, 
which captures directional curvature information beyond what is contained in the Ricci scalar.    

The remainder of this article is structured as follows. To set the stage, Sec.\ \ref{sec:found} contains a brief review of classical 
Riemannian curvature, including the related notions of parallel transport and holonomy. Motivated by quantum considerations,
we describe in Sec.\ \ref{sec:quant} the concept of the deficit angle, a standard way of quantifying 
the curvature of piecewise flat spaces, like those of Regge calculus or dynamical triangulations. 
We discuss the shortcomings of this prescription, and that of the closely related gravitational Wilson loop, in the
nonperturbative quantum theory. In Sec.\ \ref{sec:qrc}, we define the quantum Ricci curvature (QRC) and its implementation.
We summarize what is known about its behaviour on classical smooth spaces and a variety of piecewise flat spaces. 
We introduce the associated diffeomorphism-invariant curvature profile and discuss the averaging
properties of the QRC.  
Sec.\ \ref{sec:qapp} is devoted to proper quantum applications of the QRC, in 2D toy models
of Lorentzian (CDT) and Euclidean (DT) quantum gravity, and in the full 4D theory defined in terms of CDT, which
has produced important additional evidence for the de Sitter nature of its emergent geometry. 
The presence of a well-defined notion of quantum curvature in the nonperturbative quantum theory opens new
channels of investigation into the nature of quantum spacetime at the Planck scale, as we will outline briefly in Sec.\ \ref{sec:future}.
It also brings us a step closer to (semi-)classical treatments of gravity, including in the very early universe, to which we ultimately
want to connect our Planckian findings.

\section{Curvature: classical foundations}
\label{sec:found}

\begin{figure}[t]
\centerline{\scalebox{0.4}{\rotatebox{0}{\includegraphics{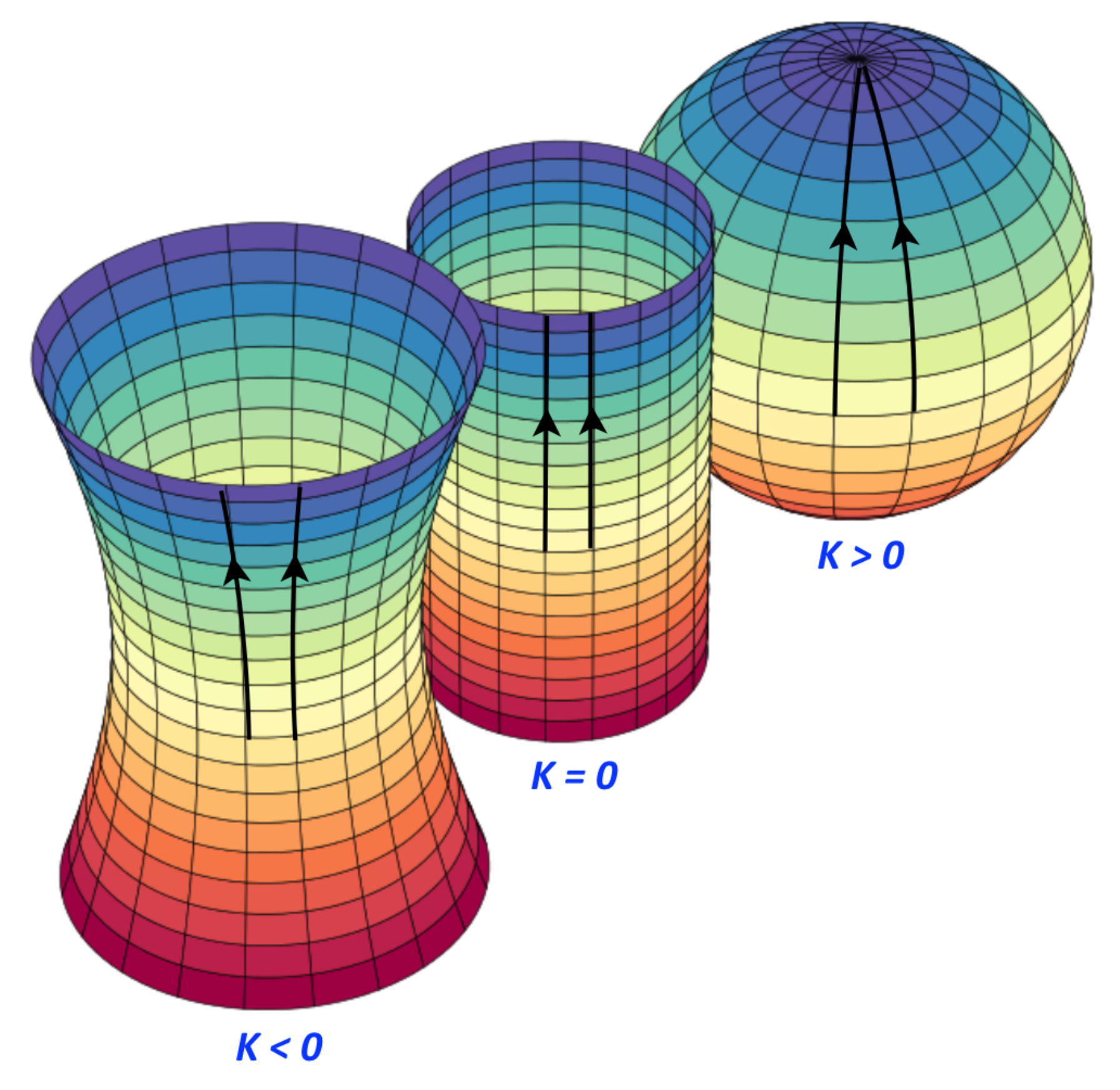}}}}
\caption{The behaviour of initially parallel geodesics depends on the Gaussian curvature $K$ of the underlying two-dimensional manifold.}
\label{fig:geodev}
\end{figure}

Before discussing possible generalizations, let us recall some aspects of classical curvature to provide background and context.
We will only consider intrinsic curvature, which is independent of any embedding of the space or spacetime under consideration. 
One way to operationally detect curvature is by following initially parallel geodesics. On a flat space, they will remain parallel, while
in the presence of curvature they can move toward or away from each other (see Fig.\ \ref{fig:geodev}). The equation of geodesic deviation
describes how the relative acceleration of neighbouring geodesics is related to the Riemann tensor $R^\kappa{}_{\lambda\mu\nu}(x)$, 
see e.g.\ \cite{Carroll2004}. 

The larger the dimension $D$ of a space, the more complicated are its curvature properties: the Riemann tensor, 
which depends on the metric $g_{\mu\nu}(x)$ and its first and second derivatives, 
has $D^2(D^2\! -\! 1)/12$ independent components, that is, 1, 6 and 20 components in $D\! =\! 2$, 3 and 4 respectively.
Another potential source of complication, especially from the viewpoint of the quantum theory, is the fact that the standard 
classical description of geometry and gravity has a large degree of redundancy, associated with
the freedom to choose any coordinate system $\{ x^\mu\}$ on the underlying spacetime. The explicit functional form of the metric
$g_{\mu\nu}(x)$ and other tensorial quantities depends on this choice, and will transform nontrivially when 
performing a coordinate change. However, the physical, geometric properties of a spacetime are independent of such a coordinate
choice and invariant under a change of coordinates. Like in a gauge field theory, in General Relativity one therefore should
distinguish carefully between ``physics" and ``gauge", that is, genuine coordinate- (or diffeomorphism-)invariant behaviour and unphysical 
``coordinate effects".\footnote{Whenever talking about coordinate-invariance, we mean invariance under the diffeomorphism group
$\mathrm{Diff}(M)$ as an active group of point transformations of the manifold $M$.} 

Taking the arbitrariness of the coordinate choice into account, curvature appears to be a more physical notion than the metric,
in the sense that the former cannot be ``transformed away'': by going to the coordinate system of a freely falling observer,
a so-called local inertial frame, the metric $g_{\mu\nu}$ is transformed locally to the metric $\eta_{\mu\nu}$
of flat Minkowski space, with vanishing first derivatives,
but its curvature (second derivatives) in general cannot be made to vanish. Simi\-larly, local scalars under coordinate transformations
are {\it curvature} invariants like the Ricci scalar $R(x)$ or the Kretschmann scalar $R^{\kappa\lambda\mu\nu}R_{\kappa\lambda\mu\nu}(x)$.
It suggests that in the context of a generalized, nonclassical geometric setting, 
some form of quantum curvature may be a more physical notion than that of a ``quantum metric $\hat{g}_{\mu\nu}$''.  
This is to some extent borne out by the nonperturbative formulation based on CDT, which does not suffer from
coordinate redundancies, and allows us to introduce the quantum Ricci curvature (see Secs.\ \ref{sec:qrc} and \ref{sec:qapp} below). 

\begin{figure}[t]
\centerline{\scalebox{0.5}{\rotatebox{0}{\includegraphics{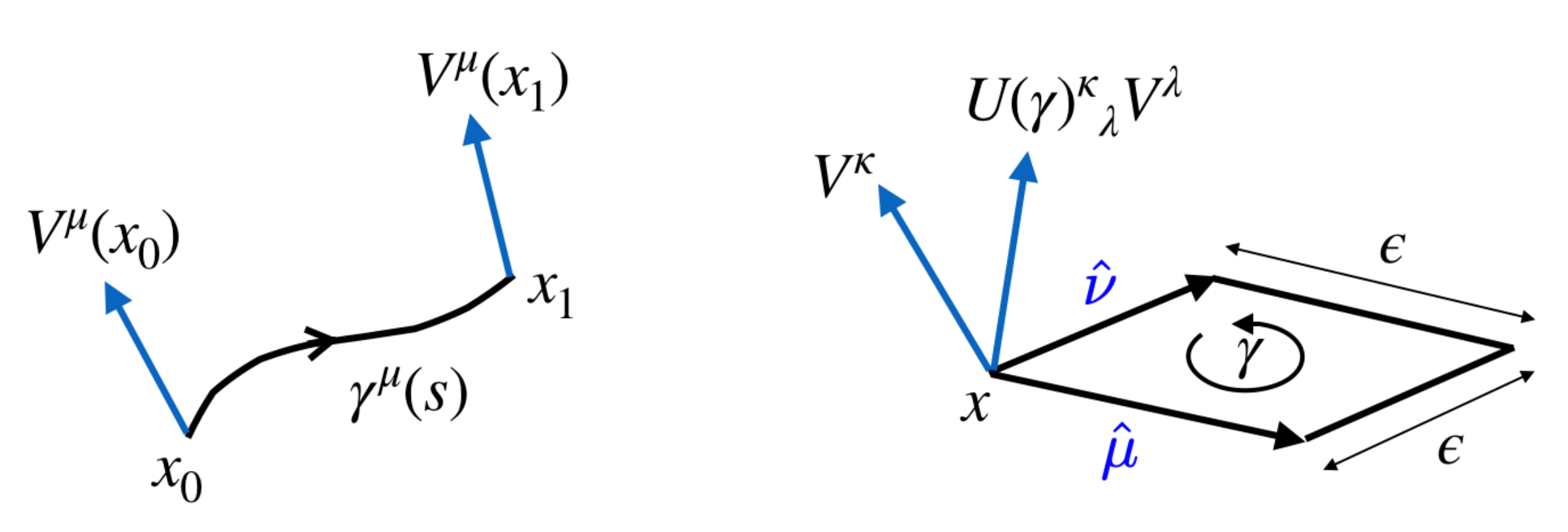}}}}
\caption{Left: parallel transport of a vector $V$ along an open path $\gamma$. Right: parallel transport of a vector $V$ around an
infinitesimal square path $\gamma$ spanned by two unit vectors in the $\mu$- and $\nu$-directions.}
\label{fig:parallel}
\end{figure}

An alternative to geodesic deviation to extract the curvature of a metric mani\-fold $(M,g_{\mu\nu})$ is to consider the 
holonomies of infinitesimal closed curves in $M$. Given a parametrized path $\gamma(s)\! : [0,1]\!\mapsto\! M$, with $\gamma (0)\! =\! x_0$
and $\gamma(1)\! =\! x_1$, the {\it holonomy} $U(\gamma ; x_0,x_1)$ is the path-ordered exponential of the
metric-compatible connection $\Gamma^\mu_{\kappa\nu}(x)$,
\begin{equation}
U(\gamma ; x_0,x_1)={\rm P}\, {\rm e}^{-\int_0^1 ds\, \Gamma_\kappa(\gamma (s))\dot{\gamma}^\kappa(s)},\;\; 
(\Gamma_\kappa)^\mu{}_\nu:=\Gamma^\mu_{\kappa\nu},
\label{holo}
\end{equation}
which is independent of the parametrization of $\gamma$. The path-ordering prescription (``P") in eq.\ (\ref{holo}) 
is needed because of the non-abelian character of the connection $\Gamma$.
The holonomy describes the parallel transport of a vector $V^\mu (x_0)$ from the point $x_0$ to the point $x_1$ 
along the path $\gamma$ (Fig.\ \ref{fig:parallel}, left) according to
\begin{equation}
V^\mu(x_1)=U(\gamma;x_0,x_1)^\mu{}_\nu V^\nu (x_0).
\label{ptrans}
\end{equation}
\noindent In a construction that may be familiar from gauge field theory (with a $su(N)$-valued gauge connection $A_\mu(x)$
instead of the Levi-Civita connection $\Gamma_\mu(x)$), one can extract curvature information by considering 
the holonomy of an infinite\-si\-mal  closed square loop $\gamma_{[\mu\nu ]}$ of side length $\epsilon$ in the $\mu$-$\nu$-plane
with base point $x$, and expanding it in powers of $\epsilon$,
\begin{equation}
U(\gamma_{[\mu\nu]};x)\! =\!{\rm P} \exp \oint_{\gamma_{[\mu\nu ]}}\!\!\!\!\!\! (-\Gamma) 
\! =\! \mathbb{1} +\epsilon^2 R_{\mu\nu}(x)+{\cal O}(\epsilon^3)
\! =\! {\rm e}^{\epsilon^2 R_{\mu\nu}}+{\cal O}(\epsilon^3), \; (R_{\mu\nu})^\kappa{}_\lambda\! :=\! R^\kappa{}_{\lambda\mu\nu}.
\label{loop}
\end{equation}
Depending on whether the metric is Riemannian or pseudo-Riemannian, the matrix $R_{\mu\nu}(x)$ 
for fixed $\mu$, $\nu$ generates an infinitesimal $SO(D)$- or $SO(1,D\! -\! 1)$-rotation in the tangent space $T_x M$ at the base point 
$x$ of the loop $\gamma_{[\mu\nu ]}$ (Fig.\ \ref{fig:parallel}, right).  
Relation (\ref{loop}) illustrates the origin of the complexity of curvature, in the sense that the number of 
components of the Riemann tensor grows very quickly with increasing $D$: it is due to the various $D$-dimensional rotations associated with
every choice of a local bi-vector spanning an infinitesimal square loop $\gamma_{[\mu\nu ]}$ around which a vector can be
parallel-transported. Lastly, note that the group-valued rotation matrix (\ref{loop}) still
transforms nontrivially under a coordinate transformation. To reduce this coordinate dependence, one can take its trace 
to obtain a ``gravitational Wilson loop"
$W_{\gamma_{[\mu\nu]}}[\Gamma]\! =\! {\rm Tr}\, U(\gamma_{[\mu\nu]};x)$, which is a scalar with respect to linear transformations of the tangent space at $x$.

\section{Taking a quantum perspective}
\label{sec:quant}

The classical curvature constructions reviewed in the previous section rely on the presence of a differentiable manifold and the
associated tensor calculus. In addition, when solving the Einstein equations (\ref{ein}), one usually requires the metric to be at 
least twice differentiable. 
As already mentioned in the introduction, none of these features may be realized in a nonperturbative regime of quantum gravity.
The construction of quantum observables -- a key aim of any quantum theory -- must therefore rely on less regular ingredients. 
It raises the question of whether meaningful notions of curvature, or quantum implementations thereof, can be defined in such a
context at all. This issue cannot be answered straightforwardly, because it goes beyond the realm of standard perturbative
quantum field theory, where local fields and operators are generically singular, but the smooth nature of the underlying manifold
is not usually questioned.    

It is tempting to argue on physical grounds that a notion like curvature, which plays such a key role in classical 
General Relativity, ought to have {\it some} correlate in the quantum theory. Otherwise, one would have to envisage
a mechanism by which curvature merely ``emerges" macroscopically from some (sub-)Planckian spacetime
substrate. Given the highly complex nature of curvature in four dimensions, this is a challenging task.
Ultimately it is a question about the true, microscopic nature of spacetime, one of the core
issues a theory of quantum gravity is expected to settle. 

There are clearly also technical issues that will need to be addressed. Even on a smooth spacetime background, curvature is a 
second-order differential operator, which in the quantum theory will generally require regularization and renormalization. This situation is unlikely to 
improve for less regular backgrounds. An example of what one may expect are the typical spacetime configurations
that contribute to nonperturbative gravitational path integrals formulated in terms of dynamical triangulations (see Fig.\ \ref{fig:DT2Dconfig} for a 2D
example), which in 
a continuum limit can be thought of as higher-dimensional analogues of the nowhere differentiable paths 
of the path-integral quantization of a nonrelativistic particle (see e.g.\ \cite{Chaichian2001} 
for a discussion of the latter).

Further difficulties may be present if one adopts graph-like or discrete models of spacetime, as advocated
in some approaches to quantum gravity, for example, the causal set approach \cite{Surya2019}.  
Nevertheless, as we will show in Secs.\ \ref{sec:qrc} and \ref{sec:qapp} by explicit construction, a notion of
Ricci curvature can be defined and implemented successfully in specific nonperturbative quantum-gravitational settings, 
thereby answering the general existence question in the affirmative. Before delving into the details, we will in the
following two subsections discuss previous attempts at defining curvature observables in quantum gravity, involving
the notion of a deficit angle, and the gravitational Wilson loop mentioned above.
Finally, let us mention that there is ongoing research in mathematics on developing generalized notions of
curvature, which can be applied on nonsmooth metric spaces, both in Riemannian and Lorentzian signature, 
see e.g.\ \cite{Ollivier2011,Cavalletti2022}.

\subsection{Curvature as deficit angles}
\label{sec:def}

\begin{figure}[t]
\centerline{\scalebox{0.52}{\rotatebox{0}{\includegraphics{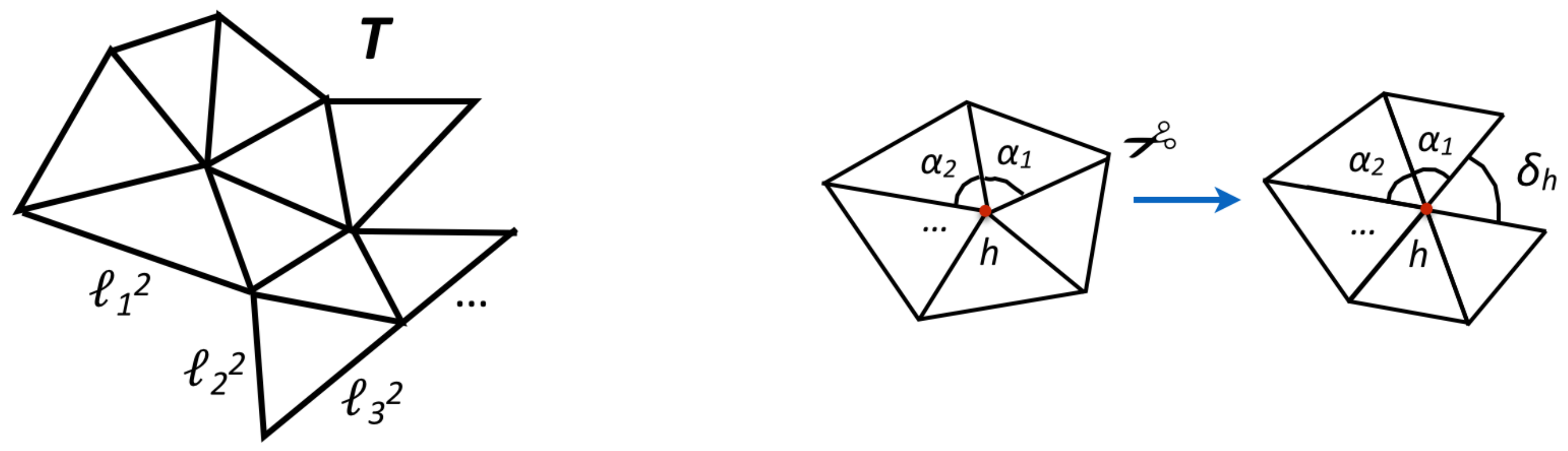}}}}
\caption{Left: two-dimensional flat triangles with edges of squared length $\ell_1^2$, $\ell_2^2$, $\ell_3^2$, ..., are glued together pairwise, 
yielding a finite triangulation $T$ with the topology of a disc. 
Right: Each of the triangles meeting at the vertex $h$ contri\-butes an angle $\alpha_i$. Cutting the triangulation open as indicated and 
putting it on a flat surface reveals a deficit angle $\delta_h$.}
\label{fig:reggetriangles}
\end{figure}

A time-honoured way to describe the spacetime dynamics of General Relativity in an approximate manner, 
without ever introducing coordinates, is {\it Regge calculus} \cite{Regge1961}. Its key
idea is to approximate metric manifolds $(M,g_{\mu\nu})$ by piecewise flat, triangulated spaces, so-called simplicial 
manifolds. In this setting, a tri\-an\-gu\-lated spacetime $T$ is characterized uniquely by the squared\footnote{The use of {\it squared} edge
lengths is important in Lorentzian signature, to distinguish between time-, space- and lightlike edges. Note that the flat, interior geometry
of a $D$-simplex is uniquely determined by its (squared) edge lengths.} edge lengths $\{\ell_i^2, i=1,2,\dots\}$ 
of its elementary building blocks, the $D$-simplices ($D$-dimensional generalizations of triangles)
and by gluing or connectivity data, which specify how the simplices are assembled into $D$-dimensional spaces 
by identifying them pairwise along matching $(D\! -\! 1)$-dimensional faces (Fig.\ \ref{fig:reggetriangles}, left). 
This method was originally devised to numerically solve the classical Einstein equations in the
absence of symmetries \cite{Williams1991,Gentle2002}. However, as has become apparent over time,
the true power of using piecewise flat spaces lies in providing a regularization for the curved spacetimes
$[g]\in{\cal G}$ that are summed (or integrated) over
in a nonperturbative version of the gravitational path integral, aka the ``sum over histories"
\begin{equation}
Z=\int_{\cal G}\! {\cal D}[g]\, {\rm e}^{i S[g]},\;\;\;\;\; {\rm with}\;\; S[g]=\frac{1}{16 \pi G_N} \int_M d^4 x\, (R-2 \Lambda).
\label{pathint}
\end{equation}
In (\ref{pathint}), $[g]$ labels a geometry, i.e.\ an equivalence class of metrics under the action of the diffeomorphism group $\mathit{Diff}(M)$,
$S[g]$ is the gravitational Einstein-Hilbert action, including a cosmological term, $G_N$ is Newton's constant and $\Lambda$ the cosmological constant. 
In CDT, the formal expression (\ref{pathint}) is substituted by a concrete prescription, which defines the nonperturbative path integral $Z$
as a continuum limit of a regularized sum over causal triangulations,
\begin{equation}
Z=\!\! \int {\cal D} [g]\; \mathrm{e}^{\, iS[g]} \;\;\; \xrightarrow{CDT} \;\;\; Z\! =\!\lim_{N_{4}\rightarrow\infty}\; \sum_{{\mathrm{causal}}\, T} 
\frac{1}{C(T)}\; \mathrm{e}^{\, iS[T]},
\label{pathcdt}
\end{equation}
where $C(T)$ is the number of elements in the automorphism group of the triangulation $T$, $S[T]$ is the simplicial implementation of
the Einstein-Hilbert action \cite{Regge1961}, and $N_4$ is the upper cutoff on the number of 
four-dimensional simplicial
building blocks contained in $T$. 
\begin{figure}[t]
\centerline{\scalebox{0.52}{\rotatebox{0}{\includegraphics{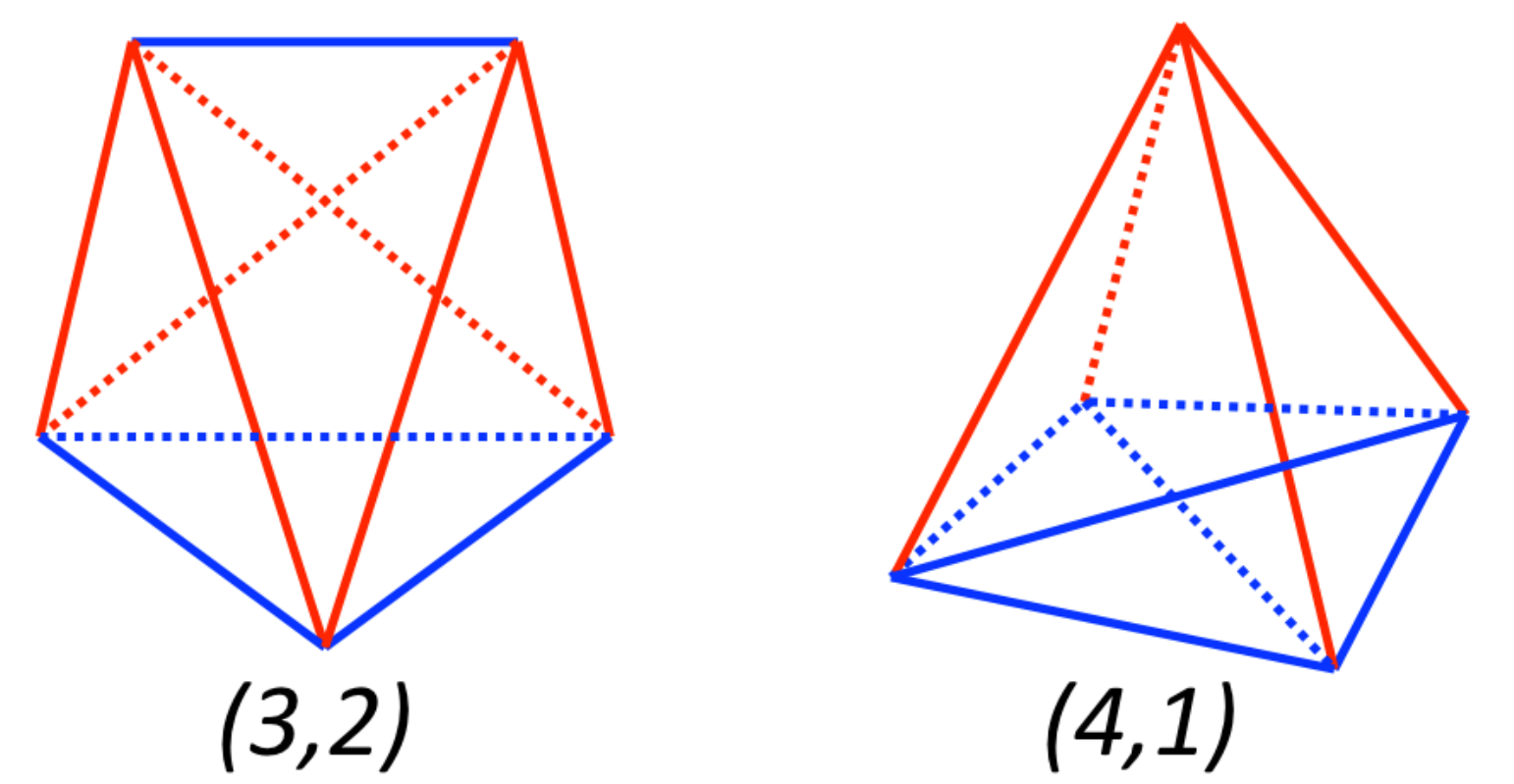}}}}
\caption{The two building blocks of 4D CDT are four-dimensional Minkowskian simplices with spacelike edges of squared length $a^2$ (blue) 
and timelike edges of squared length $-\alpha a^2$ (red): the $(3,2)$-simplex
(left) and the $(4,1)$-simplex (right).}
\label{fig:foursimp}
\end{figure}
By construction, a causal triangulation $T$ contributing to (\ref{pathcdt}) contains only two types of edges,
spacelike edges of squared length $\ell_s^2\! =\! a^2$, and timelike edges of squared length $\ell_t^2\! =\! -\alpha a^2$, for some fixed $\alpha\! >\! 0$.
There are (up to time-orientation) two geometrically distinct flat building blocks in 4D CDT (see Fig.\ \ref{fig:foursimp}), the $(3,2)$-simplex and
the $(4,1)$-simplex. The notation $(m,n)$ refers to the ``stacked'' structure of CDT geometries, where each vertex lies in a
three-dimensional spatial triangulation of fixed integer proper (lattice) time. A $(3,2)$-simplex has three vertices with time label $t$ and two
vertices with time label $t\! +\! 1$ (or the other way round), while a $(4,1)$-simplex has four vertices with time label $t$ and one vertex with
time label $t\! +\! 1$ (or the other way round).
One key advantage of representing geometries by piecewise flat spaces in the implementation by CDT
is the absence of (coordinate) redundancies in $Z$, thereby avoiding any technical complications related 
to gauge-fixing (see \cite{review1,review2,Loll2022} for further technical details and references). 

Important for our purposes is how curvature is encoded in triangulations of Regge-type. Like Regge calculus itself, this works
in the same way for any dimension $D$ and metric signature, where appropriate attention should be paid to the nature 
of Minkowskian angles when the signature is indefinite \cite{Sorkin1974}. Looking at a simple example of a triangulation, like that 
depicted in Fig.\ \ref{fig:reggetriangles}, left, for $D\! =\! 2$ and positive definite signature,
the presence of curvature may not be obvious, since a planar drawing cannot in general 
represent the edge lengths faithfully. Nevertheless, curvature is present at any vertex of such a triangulation where the angles
of the triangles sharing it do not sum up to $2\pi$, leading to a nonvanishing positive or negative {\it deficit angle}. 
For general dimension $D$, the deficit angle $\delta_h$
associated with a $(D\! -\! 2)$-dimensional subsimplex (``hinge") of a triangulation is defined by
\begin{equation}
\delta_h= 2\pi -\!\!\sum_{i|_{\Delta_i\supset h}}\!\!\alpha_i,
\label{deficit}
\end{equation}
where the sum is over all dihedral angles $\alpha_i$ at the hinge $h$ of the $D$-simplices $\Delta_i$ sharing $h$.
The situation in $D\! =\! 2$ and for a positive deficit angle $\delta_h$ is illustrated in Fig.\ \ref{fig:reggetriangles}, right. In two
dimensions, the deficit angle (\ref{deficit}) is a direct measure of the Gaussian curvature associated in a
distributional manner with the vertex $h$. This can be seen by parallel-transporting a vector along a closed loop $\gamma$ in the triangulation, which is
well-defined as long as $\gamma$ does not cross or touch the vertex $h$ (or any other vertex). 
Let us for simplicity assume that the loop lies inside the disc of triangles sharing $h$ and is neither self-intersecting
nor self-touching. Then, whenever $\gamma$ does not contain $h$ in its interior, the holonomy is the two-dimensional unit matrix, 
$U(\gamma;x)\! =\! \mathbb{1}$, for
any base point $x$, reflecting the flatness of the geometry enclosed by the loop. By contrast, if $h$ lies inside $\gamma$, 
the holonomy will be a $SO(2)$-rotation matrix corresponding to the angle $\delta\!\! \mod 2\pi$, for a suitably chosen orientation of $\gamma$. 
More precisely, the dimensionless number $\delta_h$ represents Gaussian curvature (with dimension of inverse length-squared)
integrated over a two-dimensional area 
element associated with the vertex $h$. A possible choice is given by the area of the polygon dual to $h$, whose corners are the
barycentres of the triangles sharing $h$. 

\begin{figure}[t]
\centerline{\scalebox{0.55}{\rotatebox{0}{\includegraphics{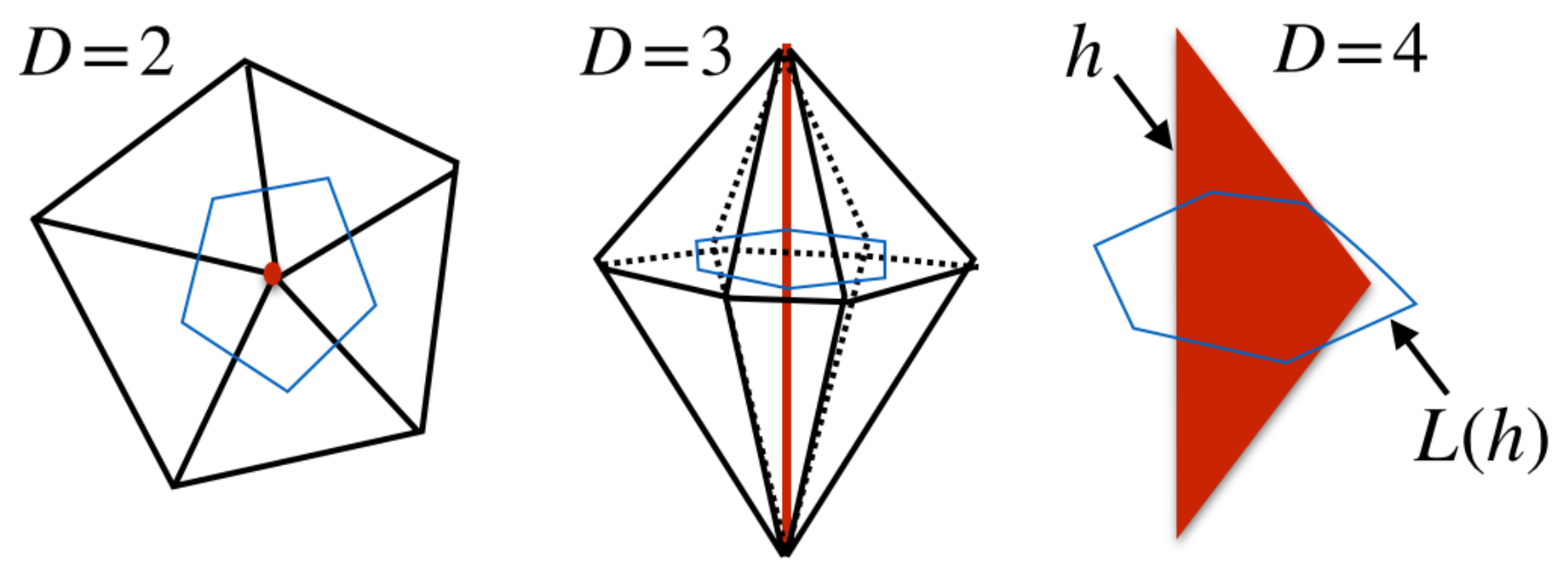}}}}
\caption{Curvature in a $D$-dimensional simplicial manifold is concentrated at hinges $h$ of dimension $D\! -\! 2$:  vertices in $D\! =\! 2$,
edges in $D\! =\! 3$, and triangles in $D\! =\! 4$ (in red). Its magnitude is given by the deficit angle $\delta_h$ at $h$, eq.\ (\ref{deficit}), and can
be extracted by considering the holonomy of a small loop $L(h)$ around $h$ (in blue).}
\label{fig:holonomy}
\end{figure}

In an analogous manner, the curvature of a higher-dimensional simplicial mani\-fold is associated with the deficit angles
at its $(D\! -\! 2)$-dimensional hinges $h$ according to expression (\ref{deficit}). The curvature at a given $h$ is directly related 
to the holonomy of a small loop $L(h)$ encircling $h$ 
in a two-dimensional area perpendicular to the hinge (see Fig.\ \ref{fig:holonomy}). An important
quantity in gravity is the integrated scalar curvature, which is part of the Einstein-Hilbert action $S[g]$ in 
(\ref{pathint}), and a particular instance of a diffeomorphism-invariant observable $\cal O$. 
Following Regge, its simplicial analogue in any dimension $D\! \geq\! 2$ 
is a weighted sum of the deficit angles of all hinges in the triangulation $T$,
\begin{equation}
\frac{1}{2}  \int_M d^Dx\sqrt{|\det g|}\, {}^{(D)\!}R \;\;  \xrightarrow[\text{}]{\text{Regge}}\;\; \sum_{h\in T}{\rm vol}(h)\, \delta_h,
\label{curvtot}
\end{equation}
where ${}^{(D)\!}R$ denotes the $D$-dimensional Ricci scalar and ${\rm vol}(h)$ the $(D\! -\! 2)$-dimen\-sio\-nal volume of the hinge $h$.
As already mentioned, Regge calculus is an approximation method, and therefore subject to the usual discretization ambiguities, for example,
how local volumes and curvatures are associated with particular subsimplices of a triangulation (see e.g.\ \cite{McDonald2008}).  
It should not be mistaken as an exact representation of General Relativity. 
In what way its expressions and equations converge to those of the classical theory has been investigated in some detail. However, we
will not discuss these specifics here since they are not directly relevant for our quantum considerations below.
 
\subsection{Deficit angle curvature in dynamical triangulations}
\label{sec:defdt}

In preparation for the quantum implementation of curvature, we will consider next the deficit angle prescription of the
previous subsection for the special case of the triangulated spacetimes that appear in the gravitational path integral of CDT. 
From now on, our discussion will concentrate on spacetimes of positive definite signature, which is the case relevant for
the analytically continued path integral of the CDT formulation. Recall that CDT quantum gravity possesses an explicit and well-defined 
``Wick rotation", given by the analytic continuation $\alpha\! \rightarrow\! -\alpha$ in the lower complex $\alpha$-plane
of the parameter $\alpha$ that characterized the length $\ell_t$ of timelike edges, as introduced below eq.\ (\ref{pathcdt}).
This analytic continuation maps the regularized CDT version of the Lorentzian path integral (\ref{pathcdt}) to
a corresponding Euclidean path integral
\begin{equation}
Z^{\rm eu}\! =\!\lim_{N_{4}\rightarrow\infty}\; \sum_{{\mathit{causal}}\, T} 
\frac{1}{C(T)}\; \mathrm{e}^{ -S^{\rm eu}[T]},
\label{patheu}
\end{equation}
where $T$ labels Wick-rotated causal, piecewise flat triangulations, and $S^{\rm eu}$ is the Euclidean Regge action. 
This analytical continuation is needed to explicitly evaluate the sum over geometries $Z$ and the expectation values
\begin{equation}
\langle {\cal O}\rangle =\frac{1}{Z} \sum_{{\mathit{causal}}\, T} 
\frac{1}{C(T)}\; {\cal O}[T]\;\mathrm{e}^{ -S[T]},
\label{expval}
\end{equation}
of observables $\cal O$ \cite{Ambjorn2001,review1,review2}, where in (\ref{expval}) and in what follows we have dropped the explicit 
superscript ``eu'' for
``Euclidean".\footnote{As was demonstrated first in two spacetime dimensions \cite{Ambjorn1998}, 
working with an analytically continued Lorentzian CDT path integral yields continuum results inequivalent to 
those of Euclidean quantum gravity in terms of dynamical triangulations (DT), whose starting point is a sum over
Riemannian spaces, and which does not possess a natural notion of Wick rotation.} 

As already mentioned, a distinguishing feature of the piecewise flat geometries contributing to the CDT path integral is the fact that 
their lattice edges come in two types of different length,
depending on whether an edge was time- or spacelike before the Wick rotation. This is different from
Euclidean DT, where there is only one possible edge length, and all triangulations are therefore equilateral. 
Both situations can be viewed as particular instances of the simplicial manifolds considered in Regge calculus, which amongst other things 
allows us to use the Regge prescription (\ref{curvtot}) for the integrated scalar curvature. To simplify the analysis of this quantity, we will 
evaluate it on an equilateral triangulation; doing the same for a non-equilateral CDT configuration is slightly more involved, but does not
alter the substance of the argument. 

For an equilateral $D$-dimensional simplex all dihedral angles, that is, all interior angles $\alpha_D$ between pairs of adjacent 
$(D\! -\! 1)$-dimensional faces, have the same value $\alpha_D\! =\! \arccos 1/D$. As a consequence, the deficit angle $\delta_h$
at a $(D\! -\! 2)$-dimensional hinge $h$ in an equilateral simplicial manifold $T$ can take a discrete set of values 
\begin{equation}
\delta_h=2\pi-c_h\alpha_D,
\label{deficitdt}
\end{equation}
where the coordination number $c_h\! =\! 3,4,5,\dots$ is the number of $D$-simplices\footnote{The fact that $T$ is 
a simplicial manifold implies a lower bound of 3 for the number $c_h$ of $D$-simplices allowed to meet at a hinge.}
sharing $h$.
For the simplest nontrivial case $D\! =\! 2$, one obtains 
\begin{equation}
\delta_h = 2\pi - c_h\pi/3\;\;\;\;\;\;\;\; {\rm (D=2)}.
\label{deficit2d}
\end{equation}
The two-dimensional case is special, in the sense that Regge's deficit angle prescription (\ref{curvtot}) for the total curvature,
with $\delta_h$ given by eq.\ (\ref{deficit2d}), satisfies an exact identity. Recall that the integrated scalar
curvature of a smooth geometry $(M,g_{\mu\nu})$ on a two-dimensional compact manifold $M$ without boundary satisfies
the so-called Gauss-Bonnet theorem
\begin{equation}
\frac{1}{2}\int_M d^2 x\sqrt{g}\, R=2\pi\chi (M),
\label{gauss}
\end{equation}
where $\chi (M)$ denotes the Euler characteristic of $M$, a topological invariant. Since for a vertex $h$ we have ${\rm vol}(h)=1$,
the total deficit angle curvature of a two-dimensional equilateral triangulation is given by
\begin{equation}
\sum_{h\in T} \delta_h=\sum_{h\in T} (2\pi\! -\! c_h \pi/3)=2\pi N_0\! -\! \pi N_2= 2\pi (N_0\! -\! N_1\! +\! N_2)=2\pi\chi (M),
\label{totaldac}
\end{equation}
where $N_0$, $N_1$ and $N_2$ are the numbers of vertices, edges and triangles in $T$ respectively, and we have used
various identities satisfied by these counting variables. It follows that the simplicial geometries, which we introduced as
approximate representations of smooth curved spaces, satisfy the same Gauss-Bonnet formula, if their curvature is expressed in
terms of deficit angles. As we will argue later, from the point of view of the nonperturbative quantum theory this behaviour appears to be a curiosity
particular to two dimensions, which does not necessarily imply that the deficit angle curvature is distinguished or preferred 
on physical grounds.

However, the main problem appears when we go to higher dimensions and 
consider the total deficit angle curvature, i.e.\ the expression on the right-hand side of (\ref{curvtot}) 
in the continuum limit of the (C)DT path integral, which is given by the combined limit $N_D\!\rightarrow\!\infty$ and $a\!\rightarrow\! 0$, while keeping
the physical, dimensionful $D$-volume $V_D\! =\! a^D N_D$ fixed. 
The dimensionful lattice spacing $a$ sets the scale of the edge or link length and plays the role of an ultraviolet length cutoff.
To understand the behaviour of the total curvature of a typical geometry in this limit, we re-express (\ref{curvtot}) in powers of $N_D$, disregarding
con\-stant terms, which yields
\begin{equation}
\sum_{h\in T}{\rm vol}(h)\, \delta_h \sim N_{D-2}\, {\rm vol}(h)\,\bar{\delta}_h \sim N_D\, (N_D^{-1/D})^{D-2}\, \bar{\delta}_h  = N_D^{2/D}\, \bar{\delta}_h,
\label{esti}
\end{equation}
where $\bar{\delta}_h$ denotes the average deficit angle, and the second proportionality follows from $N_{D-2}\!\sim\! N_D$ and  
the fact that the hinge volume scales like a $(D\! -\! 2)$-volume. For $D\! =\! 2$, the final expression in (\ref{esti}) is $N_2 \bar{\delta}_h\! =
\! N_2\, 2\pi\chi/N_0\!\sim\! const$, a finite constant, but for $D\! =\! 4$ one obtains $N_4^{1/2} \bar{\delta}_h$, which in the limit $N_4\!\rightarrow\!\infty$
diverges to infinity, because the average deficit angle is (numerically) observed to not scale to zero in this limit \cite{nonscale}. 

From a quantum field-theoretic 
point of view this divergence is not necessarily pathological, but means that the ``na\"ive'' deficit angle curvature needs to be renormalized to
become physically meaningful. However, there is currently no suggestion for how to do this. In the absence of a physically well-motivated renormalization
prescription, we will define an alternative measure of curvature in Sec.\ \ref{sec:qrc} below. It not only turns out to be better behaved in the continuum limit, but is
a direction-dependent Ricci curvature, going beyond the scalar character of (\ref{curvtot}). Before introducing this so-called \textit{quantum Ricci curvature},
we will briefly return to the gravitational Wilson loop as another potential measure of curvature, which is closely related to the deficit angle.

\subsection{Curvature from Wilson loops}
\label{sec:wilson}

We saw in Sec.\ \ref{sec:found} that the holonomy of a closed loop in a manifold $(M,g_{\mu\nu})$ contains information about its curvature, 
which appeared explicitly in the expansion (\ref{loop}) for the holonomy $U(\gamma_{[\mu\nu]};x)$ of an infinitesimal square loop in the $\mu$-$\nu$-plane. 
In search of curvature observables for the quantum theory, one may therefore consider the nonlocal gravitational Wilson loop 
\begin{equation}
W_\gamma [\Gamma]={\rm Tr}\, U(\gamma)
\label{wloop}
\end{equation}
associated with a closed loop $\gamma$ in $M$, which was already mentioned at the end of Sec.\ \ref{sec:found}.
Note that due to the cyclicity of the trace, $W_\gamma$ no longer depends on the choice of the base point $x$ of the loop $\gamma$. 
The first difficulty one encounters is that, unlike what happens in gauge field theory, the Wilson loop (\ref{wloop}) of the Levi-Civita connection $\Gamma$ is
not an observable in pure gravity, for the same reason that the Ricci scalar $R(x)$ is not an observable: although both are scalar quantities 
with respect to the frame rotations induced by a diffeomorphism, such a diffeomorphism will in general move their arguments, that is, the
point $x$ or the loop $\gamma$. Neither the Ricci scalar at a point $x$ nor the Wilson loop (\ref{wloop}) of a loop $\gamma$ 
are therefore diffeomorphism-invariant quantities. 

While it is straightforward to construct a diffeomorphism-invariant observable from $R(x)$ by integrating it over the manifold $M$,
as was done on the left-hand side of (\ref{curvtot}), integrating $W_\gamma$ over all loops $\gamma$ is neither practical nor interesting,
since $W_\gamma$ is itself already a nonlocal quantity depending in a complicated way on $\gamma$ and $\Gamma$. 
Alternatively, one could try to average the Wilson loop over a class of loops that share certain invariant geometric features regarding their 
length and shape. How to do this in a background-independent manner is not an easy task. Moreover, it is not even clear in principle what type of
curvature information could be retrieved in this way from Wilson loops of non-infinitesimal size, as we will explain further below.

On the plus side, holonomies and their associated Wilson loops are easily implemented and evaluated on the piecewise flat geometries of the
CDT path integral \cite{Ambjorn2015}, unlike on smooth curved manifolds, where they generically cannot be computed at all. 
A straightforward prescription on a four-dimensional tri\-an\-gu\-lation is to use straight path segments between the 
centres of adjacent four-simplices and consider the holonomies of piecewise straight loops made from
such segments. To compute the corresponding path-ordered exponentials one must introduce a coordinate system
in each simplex at an intermediate stage of the calculation, but the dependence on this choice drops out when taking the trace. 
Working with this particular class of closed curves does not represent a loss of generality, since a continuous deformation of a loop within 
the flat parts of a triangulation (i.e.\ avoiding its curvature singularities) leaves the associated Wilson loop invariant. 

\begin{figure}[t]
\centerline{\scalebox{0.35}{\rotatebox{0}{\includegraphics{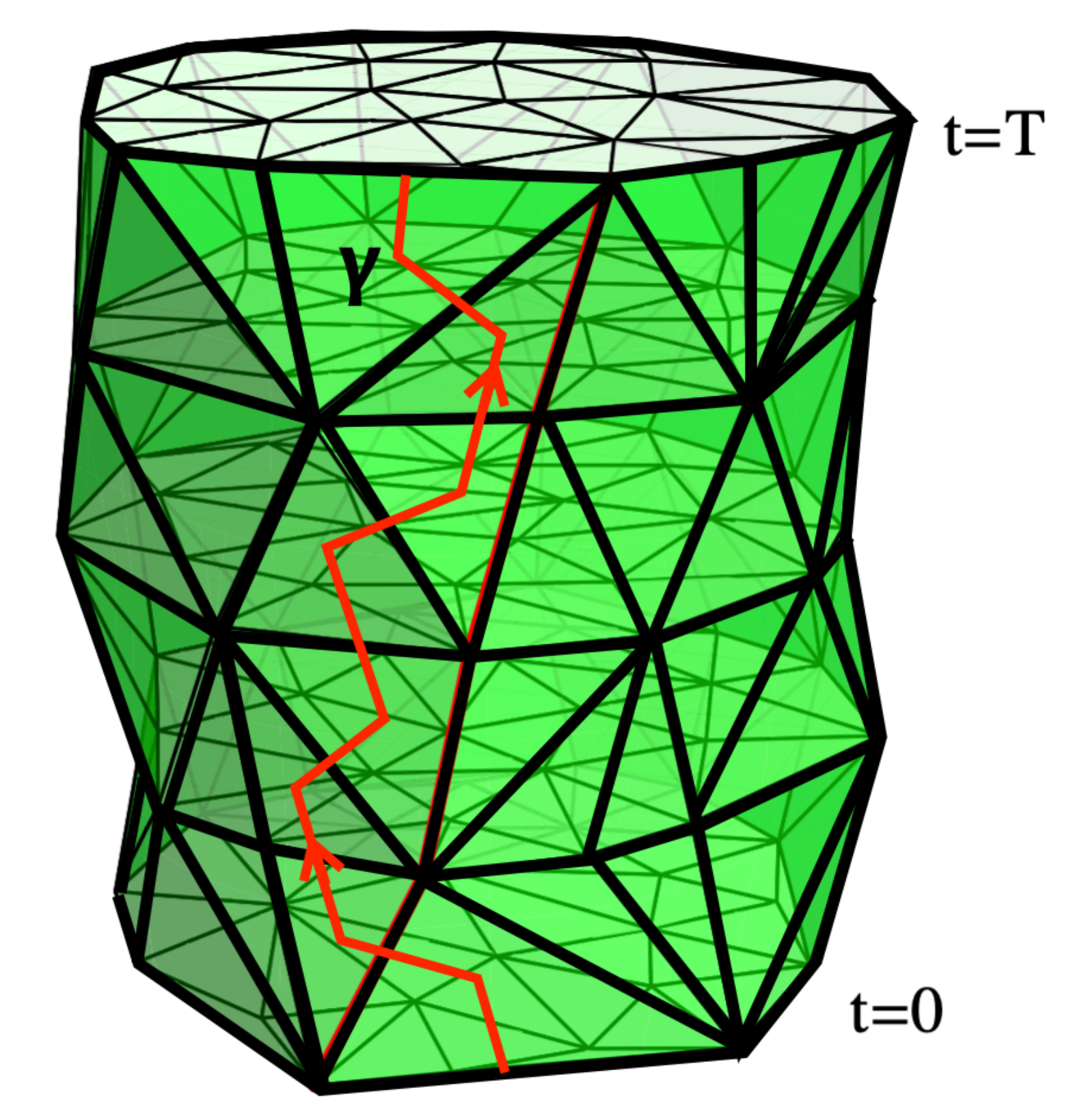}}}}
\caption{Schematic representation of the worldline of a particle, moving forward in time $t$ along a piecewise straight path in a CDT configuration, 
consisting of segments between centres of adjacent $d$-simplices. Cyclical identification of the initial and final time results in a closed loop
$\gamma$ whose corresponding Wilson loop $W_\gamma$, after summing over all loops of this type, is a quantum observable \cite{Ambjorn2015}.}
\label{fig:wilsonmassloop}
\end{figure}
The feasibility of implementing Wilson loops in four-dimensional CDT quantum gravity was demonstrated in \cite{Ambjorn2015},
which considered the expectation value of a Wilson loop whose underlying loop coincides with
the worldline of a massive particle, cyclically identified in time (Fig.\ \ref{fig:wilsonmassloop}). This is a well-defined observable, since the particle ``marks" the loop
in a diffeomorphism-invariant way. From the Wilson loop measurements one could extract the expectation value of the 
probability distribution $P(\theta_1,\theta_2)$ of two invariant angles $\theta_i\!\in\! [0,\pi ]$ characterizing $SO(4)$-elements
up to conjugation, and show that they are uniformly distributed with respect to the Haar measure on the group manifold of $SO(4)$.
It implies that, at least within the measuring accuracy of \cite{Ambjorn2015}, 
this particular Wilson loop observable does not carry any interesting curvature information (see also the remarks in \cite{review2}, Sec.\ 7.2).

Next, let us look at how a Wilson loop may in principle encode gauge-invariant information about the curvature of the underlying space.
On a smooth four-dimen\-sio\-nal Riemannian manifold,
the gravitational Wilson loop of an infinitesimal square loop $\gamma_{[\mu\nu]}$ of geodesic edge length $\varepsilon$ 
in the $\mu$-$\nu$ plane depends to lowest non\-trivial order on the entries $R^.{}_{.\mu\nu}$ of the 
Riemann curvature tensor according to
\begin{equation}
W_{\gamma_{[\mu\nu]}} = 4+\varepsilon^4 R^\kappa{}_{\lambda\mu\nu}  R^\lambda{}_{\kappa\mu\nu}+{\cal O}(\varepsilon^5),
\label{riemwilson}
\end{equation}
where the 4 comes from the trace of the unit matrix in the defining representation of $SO(4)$.
Both expressions (\ref{loop}) and (\ref{riemwilson}) reflect the fact that when we zoom in on the neighbourhood of some point $x\!\in\! M$, 
once the resolution scale $\varepsilon$ falls below
$\varepsilon\! \lesssim\! 1/\sqrt{|R(x)|}$, where $|R(x)|$ denotes the magnitude of the largest curvature at $x$, the neighbourhood will look  
ever more flat. 
However, such a flattening is not expected when we zoom in on a quantum geometry, obtained from the continuum limit of 
the gravitational path integral (\ref{pathint}), because of its nowhere differentiable character. 
As a consequence, a general holonomy $U(\gamma)$ will not be of the form ``unit matrix plus a small
perturbation", and an analogous statement holds for the corresponding Wilson loop. 

This leads one to consider \textit{non-infinitesimal Wilson loops}, exploring the possibility that the inherent
integration along a finite loop $\gamma$ may implement some effective ``averaging out" of the short-distance quantum behaviour of the 
underlying quantum geometry. Here one encounters another difficulty, related to the non-abelian character of the holonomy of the Levi-Civita 
connection. In a classical context, \textit{if} the holonomies of closed curves take values in an abelian group, one can by a suitable
version of Stokes' theorem relate the path-ordered exponential of the connection to the exponential of a
surface integral of its curvature. However, holonomies on a general curved manifold take values in all of $SO(4)$, in which
case there is no obvious way to relate finite Wilson loops to an integrated or coarse-grained form of local curvature.\footnote{There are known 
exceptions when $(M,g_{\mu\nu})$ 
has special isometries and the loops lie inside totally geodesic surfaces \cite{flux}.} Imitating the abelian
construction leads to a gravitational version of what is sometimes called the non-abelian Stokes' theorem, which in Riemannian geometry
was already considered almost 100 years ago \cite{Schlesinger1928}.  
It is not particularly useful for our purposes since
Riemann curvature occurs only in a highly nonlocal form inside the area integral appearing in the theorem, due to the presence of 
surface ordering, a two-dimensional analogue of path ordering, which is needed because of
the noncommutative nature of the connection and its associated curvature (see e.g.\ \cite{flux} and references therein). 

Despite these obstacles, we cannot rule out that in nonperturbative quantum gravity, defined by CDT or otherwise, 
suitably chosen Wilson loops could display a semiclassical behaviour matching an expression like (\ref{loop}), allowing us to
extract an effective curvature at some scale. The other possibility is that 
observables based on Wilson loops are not useful as measures of (renormalized) curvature in this context.

\section{Quantum Ricci curvature}
\label{sec:qrc}

By contrast, the quantum Ricci curvature, introduced in \cite{Klitgaard2017}, has already been used to construct well-behaved curvature observables
in nonperturbative quantum gravity. It is based on a quasi-local rods-and-clocks construction using geodesic distances and
volumes, and so far is applicable on metric spaces with positive definite signature, in particular, the Wick-rotated geometries of
the CDT formulation. As explained in greater detail below, it involves a comparison of geodesic spheres of radius $\delta$, and thereby yields
a generalized notion of Ricci curvature associated with a coarse-graining scale $\delta$. This is convenient from a physics point of view, where one
is often interested in the behaviour of physical observables as a function of a chosen length scale. 
Although the QRC can also be implemented on classical Riemannian spaces, where it reproduces the standard Ricci curvature in the limit $\delta\!\rightarrow\! 0$,
a remarkable feature of the quantum Ricci curvature is that it does not require a smooth structure, tensor calculus or parallel transport, and can
therefore be applied on a wide range of beyond-classical geometries.  

\subsection{Construction and implementation}
\label{sec:constr}

\begin{figure}[t]
\centerline{\scalebox{0.5}{\rotatebox{0}{\includegraphics{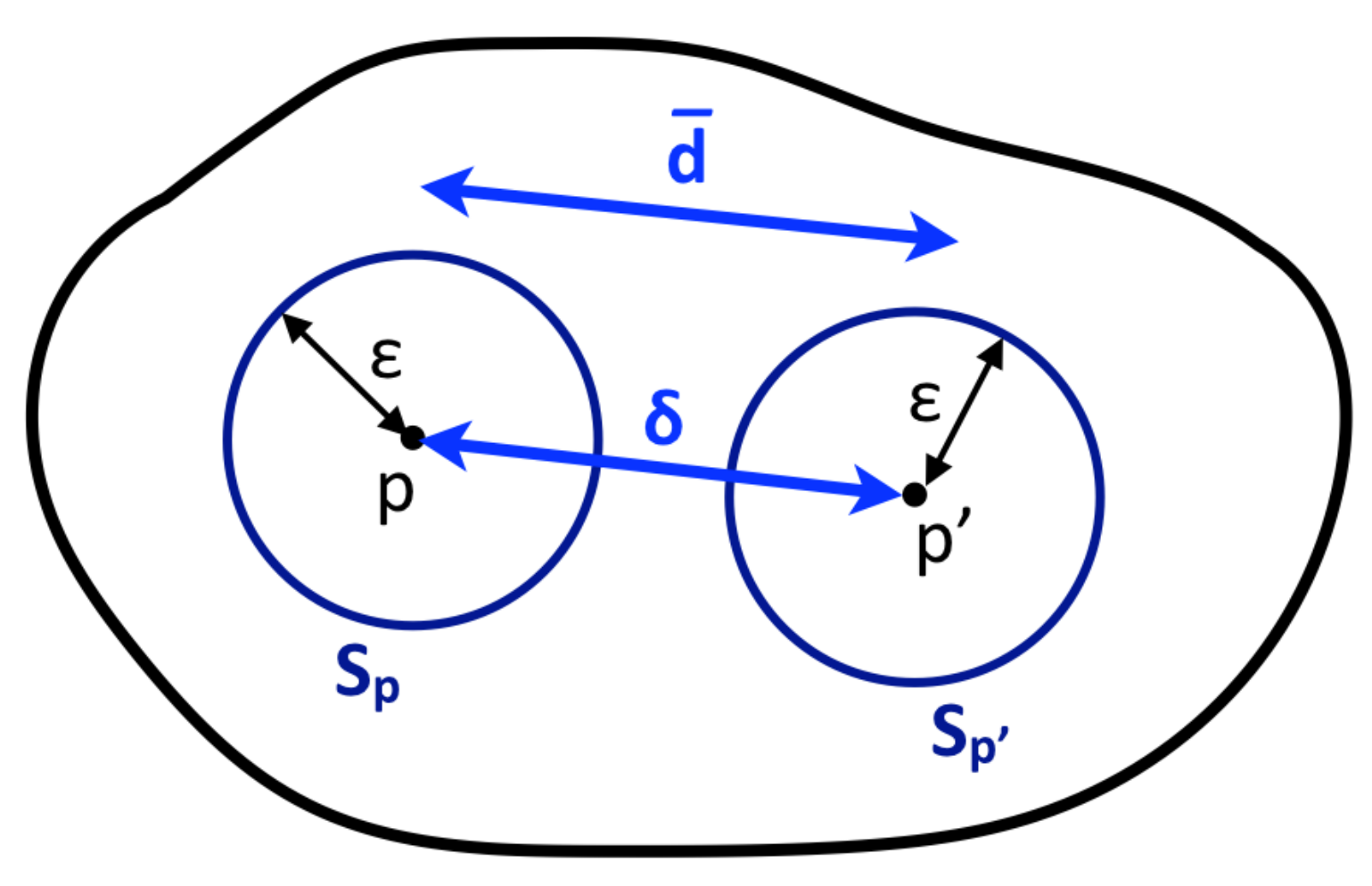}}}}
\caption{A generalized notion of Ricci curvature is obtained by comparing the distance $\bar{d}$ of two $\epsilon$-spheres $S_p$ and
$S_{p'}$ with the distance $\delta$ of their centres $p$ and $p'$.}
\label{fig:riccispheres}
\end{figure}

Given a $D$-dimensional metric measure space, the main idea underlying the QRC is to compare the distance
$\bar{d}$ of two nearby $(D\! -\! 1)$-dimensional spheres $S_p$ and $S_{p'}$ of radius $\epsilon$ (as point sets) with the distance
$\delta$ of their centres $p$ and $p'$ (Fig.\ \ref{fig:riccispheres}). The key geometric insight is then expressed by the sphere-distance criterion: 
\textit{``On a space with positive (negative) Ricci curvature, the sphere distance $\bar{d}$ is smaller (bigger) than the centre distance $\delta$."}
This observation has been used in mathematics to define a gene\-ralized notion of curvature, now known as 
Ollivier-Ricci curvature \cite{ollivier} (see also \cite{ollivier1}), which has since become a much-used tool in discrete mathematics
and graph theory \cite{jostliu,scirep}. It has also been used in graph-theoretic models for quantum gravity (see \cite{Kelly2020,Kelly2021} and
references therein).

The quantum Ricci curvature \cite{Klitgaard2017,Klitgaard2018} is inspired by this work and
has adapted it in a nontrivial way to the ensemble of piecewise flat geometries that forms the configuration space of the
nonperturbative gravitational path integral (\ref{pathcdt}) or, more precisely, its analytical continuation (\ref{patheu}). 
Firstly, in order to have only a single length scale involved, one sets $\epsilon\! =\! \delta$,
such that the two spheres overlap partially. Secondly, one uses as the distance $\bar{d}$ the 
\textit{average sphere distance}\footnote{instead of the $L^1$-transportation (or Wasserstein) distance of \cite{ollivier}, which is very expensive
computationally}, 
which in standard Riemannian continuum language is given by
\begin{equation}
\bar{d} (S_p,S_{p'}):=\frac{1}{vol(S_p)}\frac{1}{vol(S_{p'})}\int_{S_p}d^{D-1}\! q\,\sqrt{{\det h}}\int_{S_{p'}} d^{D-1}\! q'\sqrt{\det h'}
\ d_g(q,q'),
\label{spheredist}
\end{equation}
where $d_g(q,q')$ denotes the geodesic distance of the points $q$ and $q'$,
and $h$ and $h'$ are the metrics induced on the two spheres. The quasi-local \textit{quantum Ricci curvature $K_q$ at scale} $\delta$
is then defined in terms of the quotient of the average sphere distance (\ref{spheredist}) and the centre distance as
\begin{equation}
\bar{d} (S_p,S_{p'})/\delta =c_q (1-K_q(p,p')), \;\;\; \delta=d_g(p,p'),
\label{ricdefine}
\end{equation}
where $c_q$ is a non-universal $\delta$-independent constant with $0\! <\! c_q\! <\! 3$, depending on the type and dimension of 
the metric space, and $K_q$ captures the non-constant remainder. Note that $K_q$ is dimensionless by construction. It is therefore 
a curvature in units of some inverse length scale, which can be identified with $1/\delta$, at least for sufficiently small $K_q$.
When evaluating the quotient (\ref{ricdefine})
on a smooth Riemannian space for infinitesimal $\delta$, one finds that $K_q$ to lowest, quadratic order in $\delta$ 
contains $Ric(v,v)\! =\! R_{ij} v^i v^j$, the usual Ricci tensor contracted with the unit vector $v$ between $p$ and $p'$.
A com\-pu\-tation in Rie\-mann normal coordinates yields \cite{Klitgaard2018}
\begin{equation}
\frac{\bar d}{\delta}=
\left\{\begin{array}{c}
\!\!\!  1.5746+\delta^2 \left(-0.1440\, Ric(v,v)\right) +{\cal O}(\delta^3) ,\hspace{2.3cm}  D=2, \\
\!\!\! 1.6250+\delta^2 \left(-0.0612\, Ric(v,v)-0.0122\, R\right) +{\cal O}(\delta^3),\;\; D=3,\label{rnc} \\
\!\!\! 1.6524+\delta^2 \left(-0.0469\, Ric(v,v)-0.0067\, R \right) +{\cal O}(\delta^3),\;\; D=4.
\end{array}\right.
\end{equation}
\begin{figure}[t]
\centerline{\scalebox{0.45}{\rotatebox{0}{\includegraphics{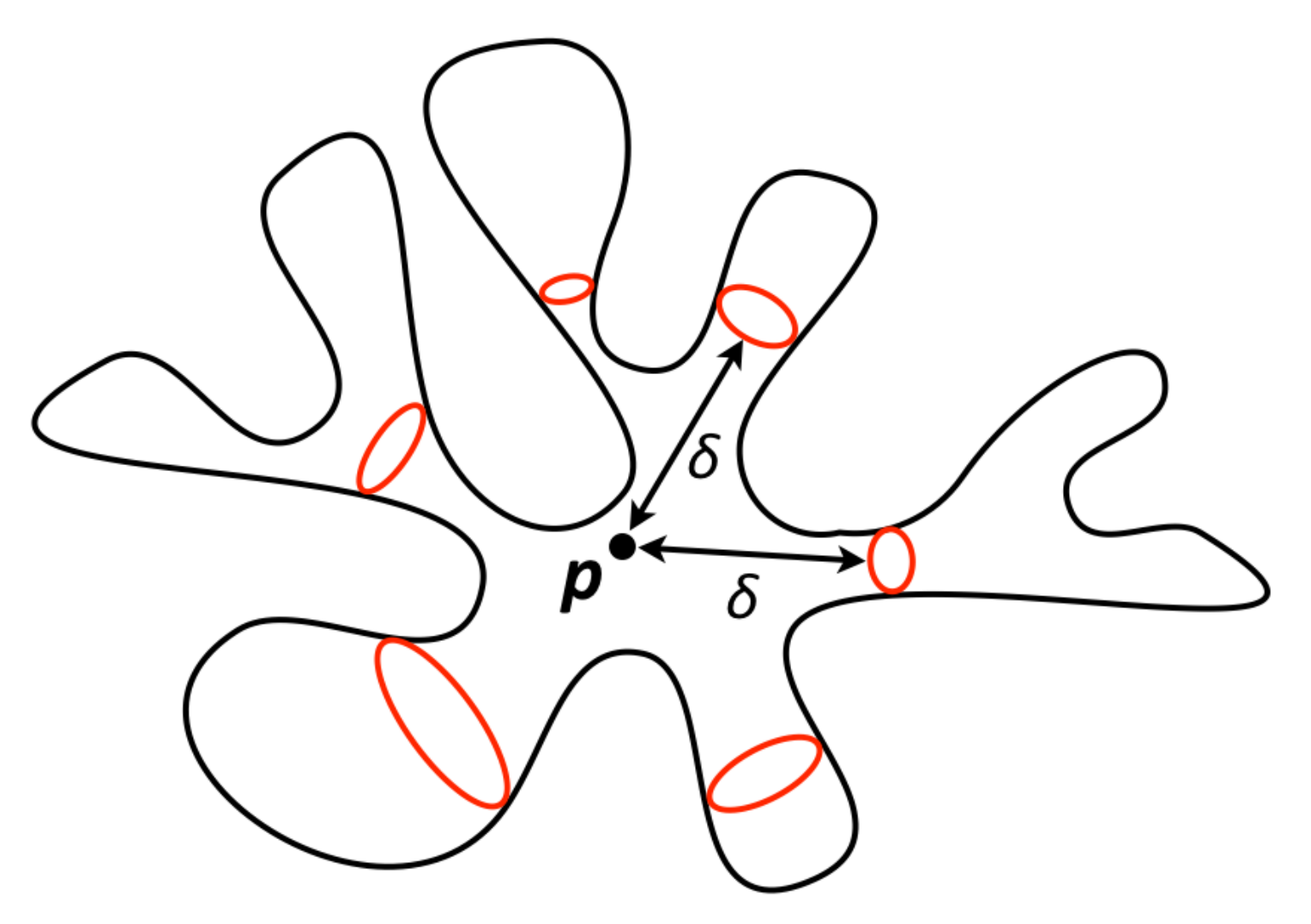}}}}
\caption{On this two-dimensional smooth manifold with $S^2$-topology, the one-dimensional ``sphere" $S_p^\delta$ with 
centre point $p$ is not a single one-sphere, but consists
of six disjoint components.}
\label{fig:branching}
\end{figure}

\vspace{-0.2cm}
\noindent Of course, the main motivation for this construction is the nonperturbative quantum theory, where the quantum Ricci curvature 
provides us with a notion of a \textit{Ricci curvature at coarse-graining scale $\delta$}, for non-infinitesimal $\delta$.
The standard notion of (discrete, geodesic) distance used on (C)DT configurations is the link distance $d(q,q')$, defined as the length
of the shortest path along lattice links between two given vertices $q$ and $q'$.\footnote{Alternatively, one may use the dual link distance, defined as the length
of the shortest path along dual lattice links between two given dual vertices. In the continuum limit, these two choices should lead
to equivalent results, but on finite lattices, one can be more convenient than the other (see Sec.\ \ref{sec:cdt4} for an example).} 
A straightforward implementation of the average sphere distance (\ref{spheredist}) on a triangulation is
\begin{equation}
\bar{d}(S_p^{\delta},S_{p'}^{\delta})=\frac{1}{N_0(S_p^{\delta})}\frac{1}{N_0(S_{p'}^{\delta})}
\sum_{q\in S_p^{\delta}} \sum_{q'\in S_{p'}^{\delta}} d(q,q'),\;\;\;\;\; d(p,p')=\delta ,
\label{spheredistlatt}
\end{equation}   
where the sphere $S_p^\delta$ 
of (integer) radius $\delta$ around a given lattice vertex $p$ is defined as the set of all vertices at link distance $\delta$ from $p$. Note
that for $\delta\! >\! 1$, this set in general does not form a topological $(D\! -\! 1)$-sphere, in the sense that the vertex set 
$S_p^\delta$ does not span a triangulated
sphere, but instead consists of several components, as illustrated by the two-dimensional example of Fig.\ \ref{fig:branching}. 
On highly nonclassical quantum geometries, this kind of branching behaviour can occur even on the smallest scales. 

In situations where one is not interested in directional information about the curvature, one can use a variant of the QRC where one lets
the two spheres of Fig.\ \ref{fig:riccispheres} coincide, i.e.\ sets $\delta\! =\! 0$ instead of $\delta\! =\! \epsilon$ to obtain a ``rotationally 
symmetric" set-up. 
The analogue of the continuum expansion (\ref{rnc}) in this case is structurally similar, but with different constants and only dependent on 
the Ricci scalar and (at higher orders) its covariant derivatives. In the construction of the curvature profile in the next subsection, we will
stick to the standard definition of the QRC and average over directions according to need.

\begin{figure}[t]
\centerline{\scalebox{0.6}{\rotatebox{0}{\includegraphics{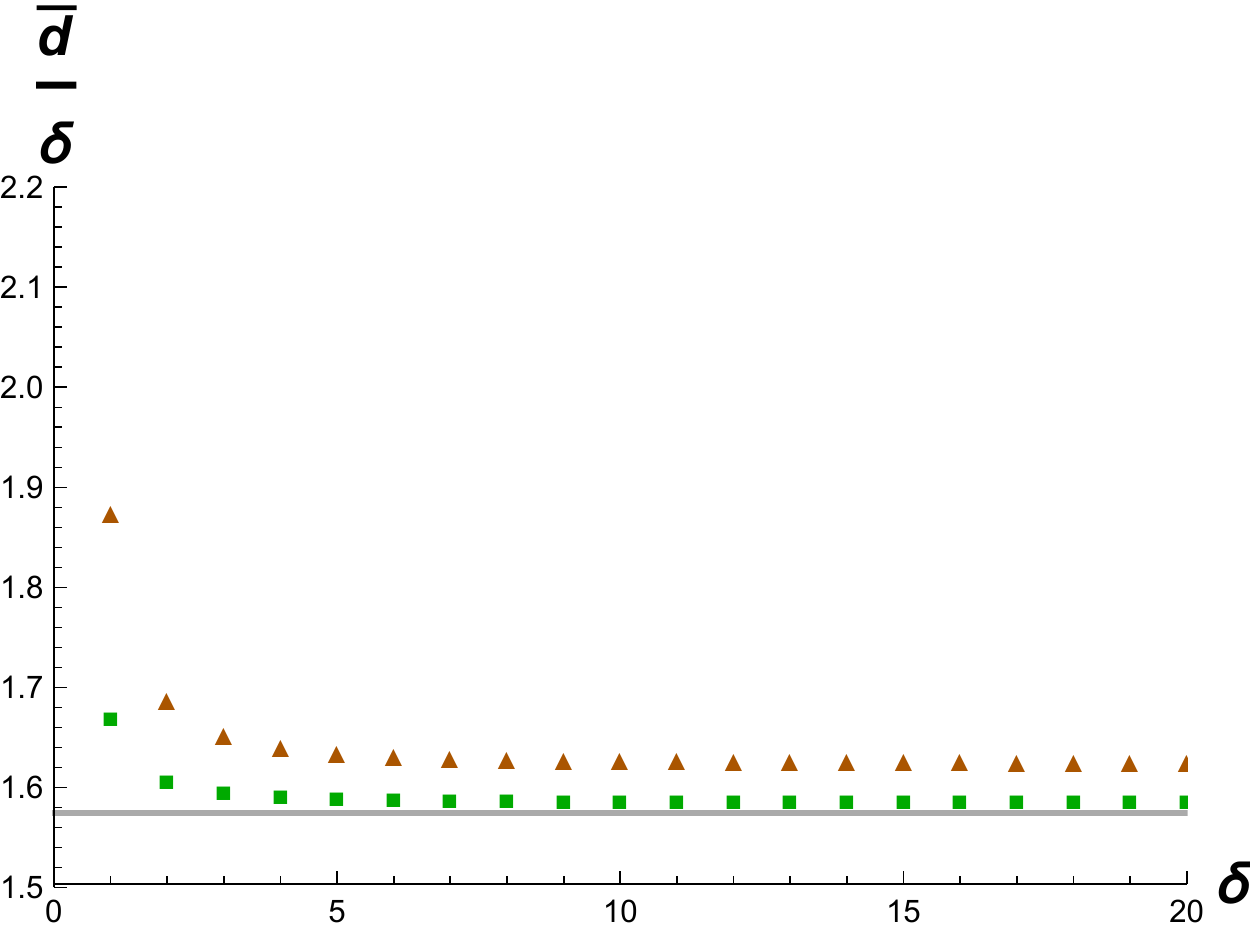}}}}
\caption{Normalized average sphere distance $\bar{d}(S^\delta_p,S^\delta_{p'})/\delta$
on regular 2D flat lattices, averaged over relative orientations of $p$ and $p'$. 
Triangles and squares mark the data points for the square and hexagonal lattices
respectively. The straight horizontal line of flat 2D continuum space is included for comparison \cite{Klitgaard2017}.}
\label{fig:2dregularav}
\end{figure}

The quantum Ricci curvature has been explored and tested on a variety of classical spaces in two and three dimensions \cite{Klitgaard2017}. 
On constantly curved Riemannian spaces, there are exact integral expressions for the average sphere distance (\ref{spheredist}). They can be
evaluated numerically and serve as benchmarks for the inter\-pretation of curvature profiles in nonperturbative applications, as we will see below.
Implementing the QRC on tesselations of flat spaces and a class of equilateral triangulations obtained from
Delaunay triangulations, which represent random approximations of constantly curved continuum spaces, gives important 
insights into the nature of lattice artefacts.
Characteristic for all lattices is the presence of a region $\delta\!\lesssim\! 5$ in link units, 
where the normalized average sphere distance $\bar{d}/\delta$ of (\ref{ricdefine}) is dominated by unphy\-si\-cal 
short-distance lattice artefacts, with an
initial ``overshoot". Furthermore, as already remarked earlier, the $\delta$-independent term $c_q$ in $\bar{d}/\delta$ is not universal, 
but depends on the type of lattice.\footnote{For a given vertex, $c_q$ can also depend
on the lattice direction, as was noted for the regular flat lattices of \cite{Klitgaard2017}. The reason is 
the highly anisotropic behaviour of the link distance on such lattices; for example, a ``sphere" $S_p^\delta$ based at a vertex $p$ of a 2D regular
square lattice is square-shaped rather than round.} 
Both of these properties are illustrated by 
Fig.\ \ref{fig:2dregularav}, which shows the behaviour of the quotient $\bar{d}/\delta$ as a function of the link distance $\delta$ on two different
lattice discretizations of two-dimensional flat space. Within measuring accuracy, both are compatible with a vanishing QRC, $K_q\! =\! 0$. 
These examples also illustrate that the constant $c_q$ of relation (\ref{ricdefine}), which in
the continuum is given by $c_q=\lim_{\delta\rightarrow 0} \bar{d}/\delta$, has to be defined appropriately on piecewise flat spaces, at some reference
point $\delta$ outside the region of lattice artefacts, typically chosen as $\delta\! =\! 5$ or $\delta\! =\! 6$.

\subsection{Curvature observables: the curvature profile}
\label{sec:prof}

Neither the average sphere distance (\ref{spheredist}), (\ref{spheredistlatt}), nor the quantum Ricci curvature extracted from it are
observables, since they still depend on a specific point pair $(p,p')$. The most straightforward way of constructing a diffeomorphism-invariant
observable is by averaging the average sphere 
distance $\bar{d}$ over all positions $p$ and $p'$ of the two circle centres, 
while keeping their distance $\delta$ fixed. Adopting again a continuum language, the average on a given compact Riemannian manifold
$(M,g_{\mu\nu})$ of the average sphere distance at the scale $\delta$ is given by
\begin{equation}
\bar{d}_{\rm av}  (\delta):=\frac{1}{Z_\delta}
\int_{M}d^D\! p\, \sqrt{\det g} \int_M d^D\! p' \, \sqrt{\det g}\;\,  \bar{d}(S_p^{\delta},S_{p'}^{\delta})\; \delta^{Dir}\! (d_g(p,p'),\delta),
\label{avdist}
\end{equation}   
where $\delta^{Dir}$ denotes the Dirac delta function and the normalization factor $Z_\delta$ is given by
\begin{equation}
Z_\delta=\int_{M}d^D\! p\, \sqrt{\det g} \int_M d^D\! p'\, \sqrt{\det g}\;\; \delta^{Dir}\! (d_g(p,p'),\delta).
\label{zdelta}
\end{equation}
Since the integration in eq.\ (\ref{avdist}) includes an averaging over directions, it only
allows us to extract an (averaged) quantum Ricci scalar $K_{\rm av}(\delta)$. 
This scalar quantity appears in the $\delta$-dependent \textit{curvature profile} \cite{Brunekreef2020}, which is defined as the quotient
\begin{equation}
\bar{d}_{\rm av}(\delta)/\delta=:c_{\rm av} (1 - K_{\rm av}(\delta)),
\label{profile}
\end{equation}
where the constant $c_{\rm av}$ is given by $c_{\rm av}\! =\!\lim_{\delta\rightarrow 0} \bar{d}_{\rm av}/\delta$.
The curvature profile is a nonlocal curvature observable characterizing a given curved manifold. 

The curvature profile is not an observable that one would naturally consider in a classical context, where one is usually
interested in resolving \textit{local} curvature properties of selected solutions to the Einstein equations. The situation is different in the
quantum theory, where one is primarily interested in the quantum geometry of the unique nonperturbative vacuum. One may
expect this quantum geometry to have some approximate global symmetries, at least on suitably coarse-grained scales, although this conjecture
still needs to be verified explicitly (see also the discussion of the
curvature profile of the de Sitter-like vacuum state of 4D CDT quantum gravity in Sec.\ \ref{sec:cdt4} below and \cite{Loll2022}). 
If such symmetries are
indeed present, the difference between local curvature properties and their space(time) averages may not be all that big.

\begin{figure}[t]
\centerline{\scalebox{0.42}{\rotatebox{0}{\includegraphics{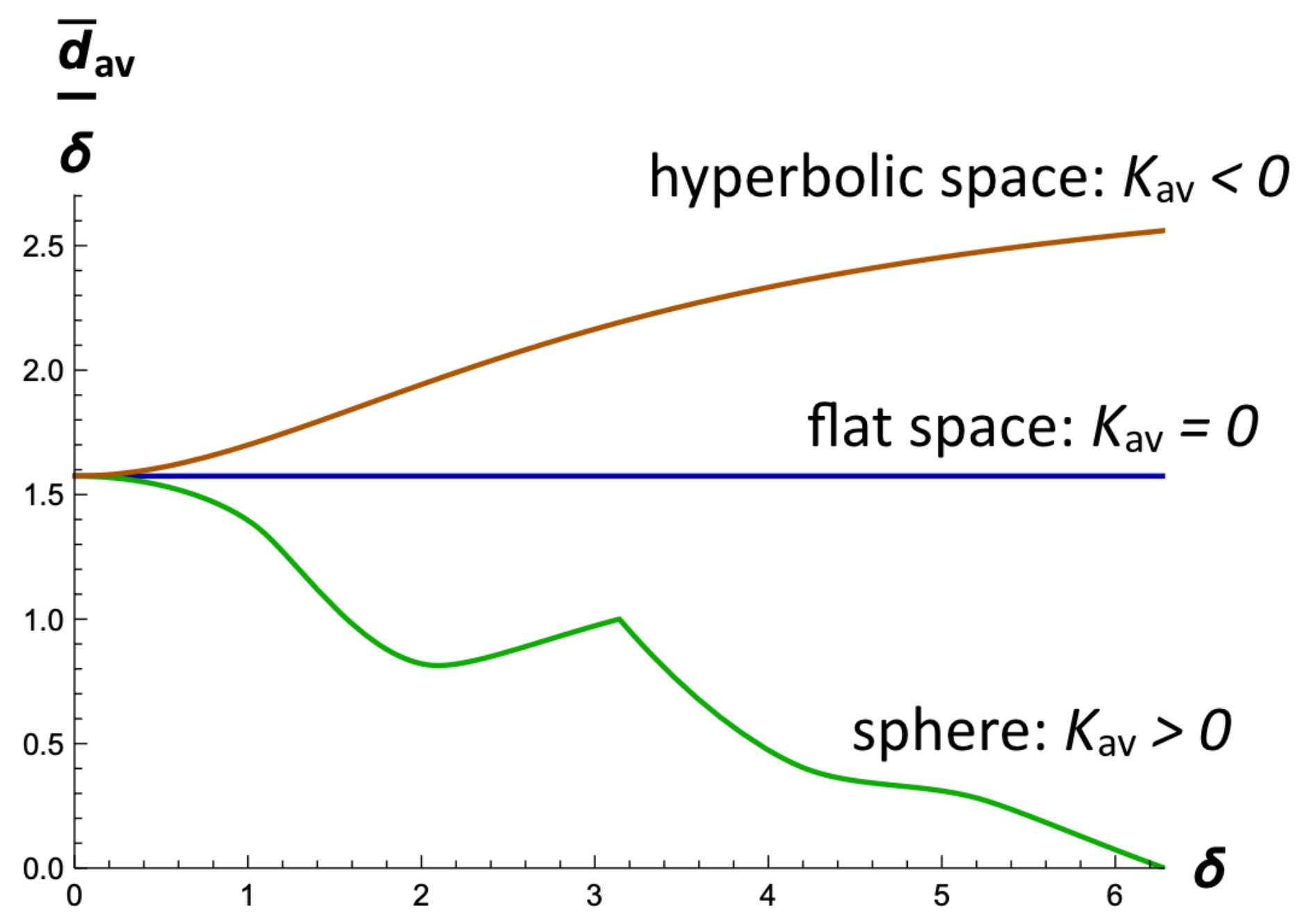}}}}
\caption{Curvature profiles $\bar{d}_{\rm av}(\delta)/\delta$ of constantly curved Riemannian spaces in $D\! =\! 2$ \cite{Klitgaard2017}.
The curvature radius has been set to 1.}
\label{fig:constcurv}
\end{figure}

In order to interpret (the expectation values of) curvature profiles obtained in the quantum theory, it will be helpful to compare them to specific
classical curvature profiles. In line with the argument of the previous paragraph, 
a natural reference point are the constantly curved differentiable manifolds of positive, negative or vanishing curvature,
that is, spherical, hyperbolic or flat spaces in any dimension $D$. Note that in these cases no averaging over $M$ is needed to obtain the
classical curvature
profile, since the average sphere distance (\ref{spheredist}) depends only on the distance $\delta$, and not on the location of $p$ and $p'$. 
The curvature profiles of spaces of constant curvature were computed in \cite{Klitgaard2017} and are shown in 
Fig.\ \ref{fig:constcurv} for $D\! =\! 2$; the analogous curvature
profiles in higher dimensions look qualitatively similar. Note that the characteristic overall shape of the curvature profile for $S^2$ has to do with its global 
properties and perfect roundness.\footnote{For example, the kink at $\delta\! =\! \pi$ corresponds to the situation where the two sphere centres 
$p$ and $p'$ are antipodes and their associated ``circles" $S_p^\delta$ and $S_{p'}^\delta$ of radius $\delta$ are degenerate and themselves 
correspond to points.
On a space that only approximates a sphere in some loose, average sense, the global features of the curvature profile may look very different.} In practice, 
when comparing with quantum measurements, only the initial regions 
of these curves are used, where $\delta$ is much smaller than the linear extension of the (quantum) space under consideration. 
The main feature of $\bar{d}_{\rm av}/\delta$ one looks for in this range is whether it increases, decreases or stays constant 
in $\delta$, corresponding to a negative, positive or vanishing average quantum Ricci scalar $K_{\rm av}$. 

The classical curvature profile (\ref{profile}) is an observable and its expectation value $\langle \bar{d}_{\rm av}(\delta)/\delta \rangle$ can 
be determined numerically in (C)DT or other formulations of quantum gravity. 
Note that in a lattice approach like CDT, all distances are given in dimensionless
lattice units, which can be converted into dimensionful units by invoking the lattice spacing $a$, introduced in Sec.\ \ref{sec:defdt}.
Recall that the dimensionless $K_{\rm av}(\delta)$ has the interpretation of a curvature in units of $1/\delta$, both classically and 
in the nonperturbative quantum theory. Assuming for the moment that $\delta$ scales canonically, that is, proportional to $a$ or,
equivalently, proportional to $N_D^{1/D}$, one can extract a dimensionful, renormalized quantum Ricci scalar 
$K^r\! (\delta_{\rm ph})$, which depends on a physical, renormalized length scale $\delta_{\rm ph}\! :=\! a\delta$, via
\begin{equation}
K_{\rm av}(\delta)=:\delta^2 a^2 K^r\! (\delta_{\rm ph})=(\delta_{\rm ph})^2 K^r\! (\delta_{\rm ph}), 
\label{qren}
\end{equation}
in the limit as $a\!\rightarrow\! 0$. In case $\delta$ scales non-canonically, the renormalized curvature will scale accordingly and also non-canonically.

\subsection{Averaging properties of the QRC}
\label{sec:ave}

Moving away from constantly curved spaces, another interesting question is to what extent the classical curvature profile is sensitive
to an inhomogeneous distribution of curvature. From the viewpoint of the quantum theory, it is desirable that any notion of curvature has good averaging
properties, unlike the deficit angle curvature, which is a local curvature defined \textit{at the cutoff scale} and
already singular at the classical level, in the sense of being concentrated on $(D\! -\! 2)$-dimensional hinges, as described in Sec.\ \ref{sec:def}.  
As we saw in Sec.\ \ref{sec:defdt}, in a continuum limit, these curvature singularities just become denser and denser, without ``averaging out''.

One class of triangulations whose curvature profiles have been studied involves 2D Delaunay triangulations 
approximating constantly curved spaces.
Delaunay triangulations are random, piecewise 
flat spaces, which by construction closely approximate their smooth counterparts in terms of their curvature properties.
One proceeds by constructing via a Poisson
disc sampling a point set $P$ in the constantly curved model space in question, with a chosen minimal distance $d_{\rm min}$ between any two of its points.
One then constructs the Delaunay triangulation that has $P$ as its vertices and subsequently sets all edge lengths to 1 
to obtain an equilateral triangulation on
which the QRC can be implemented in a straightforward way (see \cite{Klitgaard2017} for details). Up to an overshoot for small $\delta\!\lesssim\! 5$ and 
a vertical shift due to the non-universal constant $c_q$, features which were already observed in lattice representations of flat 
space (cf.\ Fig.\ \ref{fig:2dregularav}), the volume profiles for these random triangulations closely resemble those of their constantly curved 
continuum counterparts. 

\begin{figure}[t]
\centerline{\scalebox{0.42}{\rotatebox{0}{\includegraphics{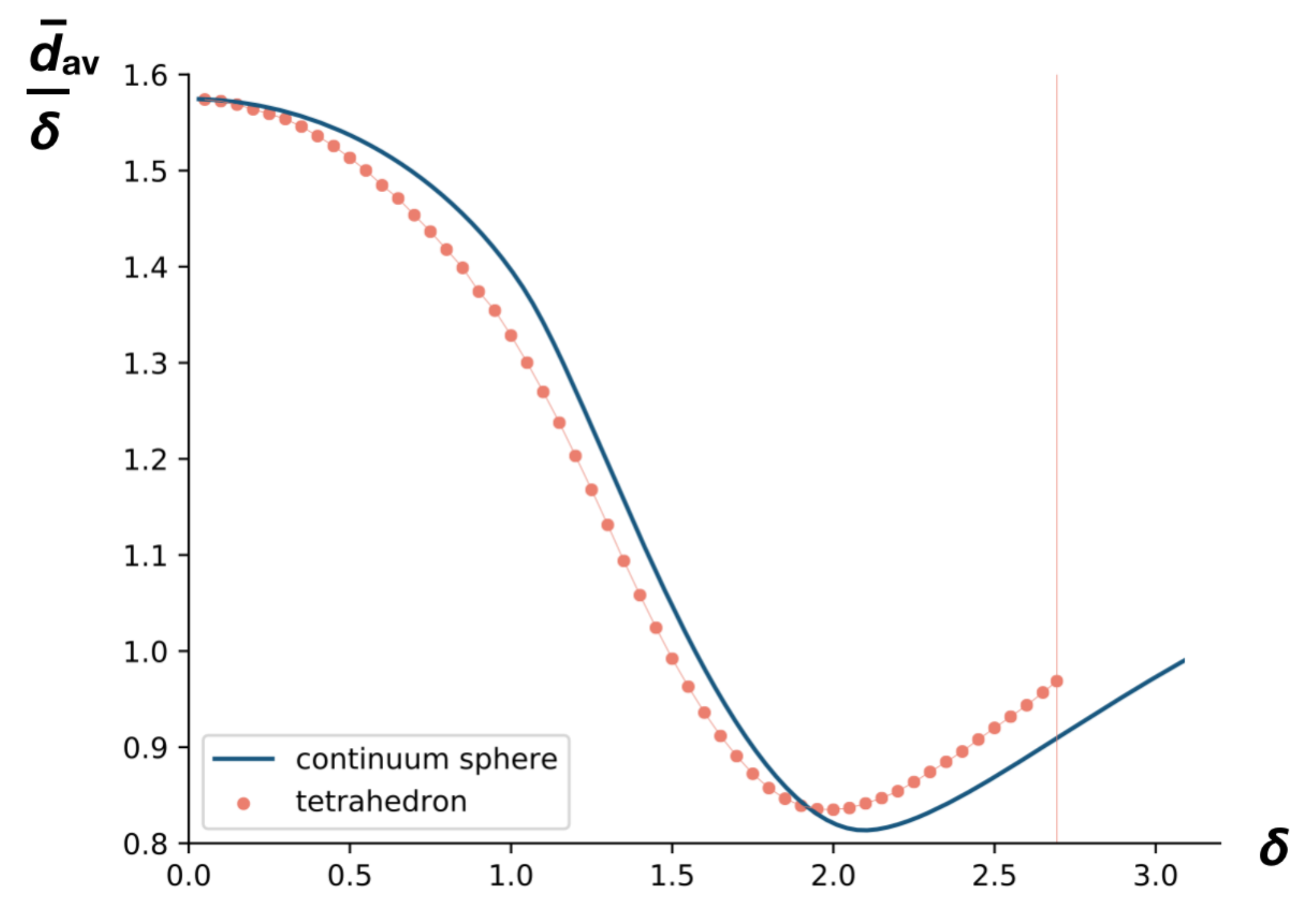}}}}
\caption{Measured curvature profile $\bar{d}_{\rm av}(\delta)/\delta$ of the surface of a tetrahedron, compared to that of a two-sphere of
the same area. The vertical line at $\delta\! =\! 2.694$ marks the edge length of the tetrahedron \cite{Brunekreef2020}.}
\label{fig:tetrahedron}
\end{figure}

Another type of classical geometry whose curvature profile has been studied is a two-dimensional space of spherical topology 
with isolated conical singularities, more specifically, a regular polygon that forms the surface of one of the Platonic solids \cite{Brunekreef2020}. 
The polygon with
the largest conical defects is the surface of a tetrahedron, which is flat except at its four corners, with a deficit angle of $\pi$ each. 
In terms of curvature (in)homogeneity, it is the extreme opposite of a round sphere, on which the same total Gaussian curvature 
($4\pi$ by the Gauss-Bonnet theorem)
is distributed completely homogeneously. In addition, a lot is known about geodesics on the tetrahedron, which allows one to construct
the geodesic circles that are key to the computation of the QRC \cite{Brunekreef2020}. 

The curvature profiles for the tetrahedral surface and for a sphere
of the same area are shown in Fig.\ \ref{fig:tetrahedron}. The reference curve for the sphere coincides with that of Fig.\ \ref{fig:constcurv} for
$\delta\! \leq\! \pi$. The two curvature profiles clearly differ, but not by very much, testifying to the averaging property of the quantum Ricci curvature.
This is a reassuring feature, but the real test of the QRC is its application in the nonperturbative quantum theory, which will be the subject 
of Sec.\ \ref{sec:qapp}. At the same time, these results emphasize the global nature of the curvature profile $\bar{d}_{\rm av}(\delta)/\delta$. 
Similar to the volume profile $V_3(\tau)$,
which denotes (the expectation value of) the spatial volume as a function of proper time in 4D CDT quantum gravity \cite{Ambjorn2007,review1,review2}, 
also the curvature profile needs to be complemented by other observables, like correlators or homogeneity measures \cite{Loll2023} 
to obtain a finer-grained understanding of the underlying quantum geometry.

\section{Quantum Ricci curvature: quantum applications}
\label{sec:qapp}

The quantum Ricci curvature, which was introduced in the previous section, gives us a new tool to characterize and quantify
\textit{quantum} geometry, including in regimes far away from classicality. It is not known a priori what kind of nontrivial quantum behaviour 
the QRC can exhibit, but there is no reason in principle why it should be any less complex than that of its classical counterpart. 
In the context of nonperturbative models based on dynamical triangulations, a natural first testing ground for the QRC are toy models in two dimensions. 
Pure gravity models in 2D, whose action consists of a (topological) Einstein-Hilbert term\footnote{Nonperturbative path integrals based on
(C)DT are defined for fixed spacetime topology.} and a cosmological-constant term, fall into two distinct 
universality classes, depending on their metric signature. CDT quantum gravity lies in the universality class of Lorentzian,
and DT quantum gravity in the universality class of Euclidean signature. Since classical gravity in two dimensions is trivial, there are no
(nontrivial) classical solutions. As a consequence, 2D quantum gravity models and the quantum geometries they generate
do not possess a nontrivial classical limit and are therefore maximally ``quantum". The curvature profiles of 2D Lorentzian and Euclidean 
quantum gravity will be described in Secs.\ \ref{sec:cdt2} and \ref{sec:dt2} below. 

However, the main motivation for the introduction of the QRC was the need to understand the curvature properties of the emergent
quantum spacetime found in full, four-dimensional quantum gravity formulated in terms of CDT \cite{Ambjorn2004,Ambjorn2005a}.
Its volume profile matches that of a classical de Sitter universe, with quantum volume fluctuations that likewise match a semiclassical treatment \cite{Ambjorn2007,Ambjorn2008}. By investigating the curvature properties of this dynamically generated quantum universe, one would like to
understand to what extent it matches the expected behaviour of the classical, constantly curved de Sitter space, and to quantify how its curvature 
deviates in a (quasi-)local sense from a perfectly homogeneous distribution. Currently available results in 4D will be summarized in
Sec.\ \ref{sec:cdt4}.

\begin{figure}[t]
\vspace{-1cm}
\centerline{\scalebox{0.4}{\rotatebox{90}{\includegraphics[trim=2.5cm 1cm 2.5cm 1cm, width=0.95\textwidth, clip]{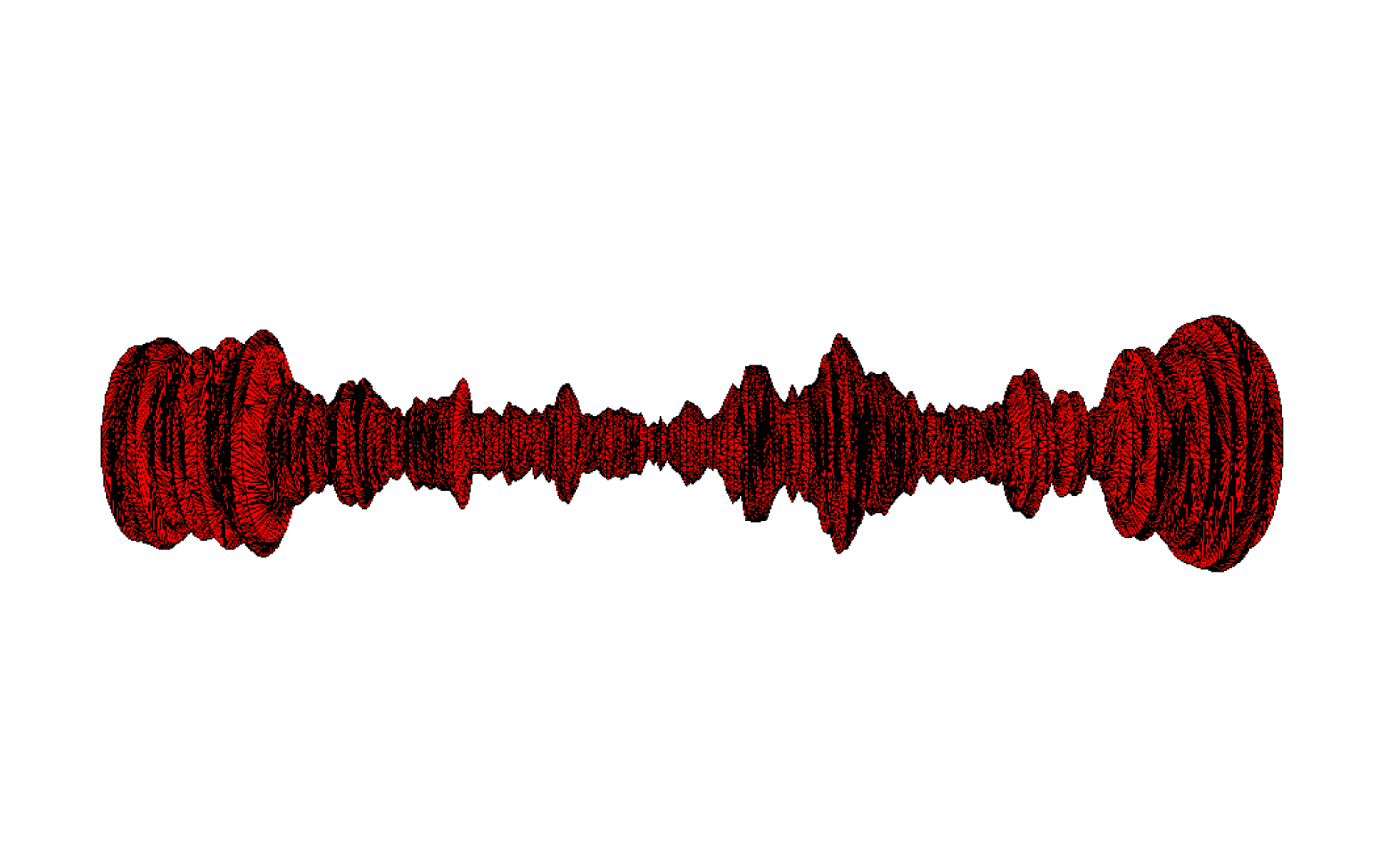}}}}
\vspace{-1cm}
\caption{Typical configuration contributing to the CDT path integral in 2D, 
for $N_2\! =\! 9408$ \cite{2dmatter}, illustrating the fluctuating volume of 
the spatial $S^1$-slices as a function of proper time (horizontal direction, cyclically identified in simulations).}
\label{fig:candlestick}
\end{figure}

\subsection{CDT quantum gravity in D\,\! =\,\! 2}
\label{sec:cdt2}

The two-dimensional version of the Wick-rotated CDT path integral (\ref{patheu}) is given by 
\begin{equation}
Z = \sum_{{\rm causal}\ T} \frac{1}{C(T)} \, {\rm e}^{-S_\lambda[T]}, \quad \quad S_\lambda[T] = \lambda\, N_2(T),
\label{path2dcdt}
\end{equation}
where the sum is taken over causal triangulations $T$ of topology $S^1\!\times\! S^1$, with compact spatial $S^1$-universes of variable length and a
compactified time direction for computational convenience. 
The Euclidean action $S_\lambda [T]$ consists of a cosmological-constant term, 
where the bare cosmological constant $\lambda$ multiplies
the discrete volume $N_2(T)$, counting the number of equilateral 2D triangles in $T$.
CDT quantum gravity in 2D was first solved analytically in \cite{Ambjorn1998}; the continuum theory has  
a spectral dimension of (at most) 2 \cite{spec2d} and a Hausdorff dimension of 2 \cite{Ambjorn1998,lessons,spec2d}. 
The fact that these dynamical dimensions happen to be equal to the topological dimension of the triangular building blocks of the
regularized path integral does not imply that the quantum geometry resembles any classical geometry and even less that it is locally flat. 
Fig.\ \ref{fig:candlestick} shows a typical member of the ensemble of 2D CDT geometries, illustrating the large fluctuations of 
the spatial volume of the universe as a function of (discrete) proper time. Clearly these universes are not individually locally flat, but it is not
clear a priori what picture will happen to the curvature when both a manifold and an ensemble average are taken into account. 

The expectation value $\langle \bar{d}_{\rm av}(\delta)/\delta \rangle$ 
of the curvature profile of the quantum geo\-metry generated in 2D CDT quantum gravity was investigated in \cite{Brunekreef2021} 
for a range $N_2\!\in\! [50k,600k]$ of spacetime volumes, for two different time extensions $\tau\! =\! 183$, 243, 
and a corresponding range $\delta\! \in\! [1,\delta_{\rm max}]$ in terms of link distance, with $\delta_{\rm max}\! =\! 30$, 40, respectively. 
As usual, the lattice simulations were performed for fixed discrete volume $N_2$, in search for a continuum (scaling) limit as $N_2\!\rightarrow\!\infty$.  
Another set of measurements with $N_2\!\in\! [100k,250k]$ and $\tau\! =\! 183$ used the dual link distance, 
with $\delta_{\rm max}\! =\! 60$. 
The maximal sphere radii $\delta_{\rm max}$ were chosen to avoid that the results are affected significantly by topological (finite-size) effects 
due to the compactified time and spatial directions.
Such an effect occurs whenever a shortest geodesic between two points $(q,q')\!\in\! S_p^\delta\!\times\! S_{p'}^\delta$ only exists because
of the compactness, and would not exist on an open patch containing the two $\delta$-spheres. The subtle part is
to control for such effects in the spatial direction, because of the strong fluctuations in the length of the spatial universe (Fig.\ \ref{fig:candlestick}).  
This was achieved in \cite{Brunekreef2021} by monitoring the relative winding numbers of geodesics between pairs $(q,q')$ of 
points.

\begin{figure}[t]
\centering
\includegraphics[width=0.7\textwidth]{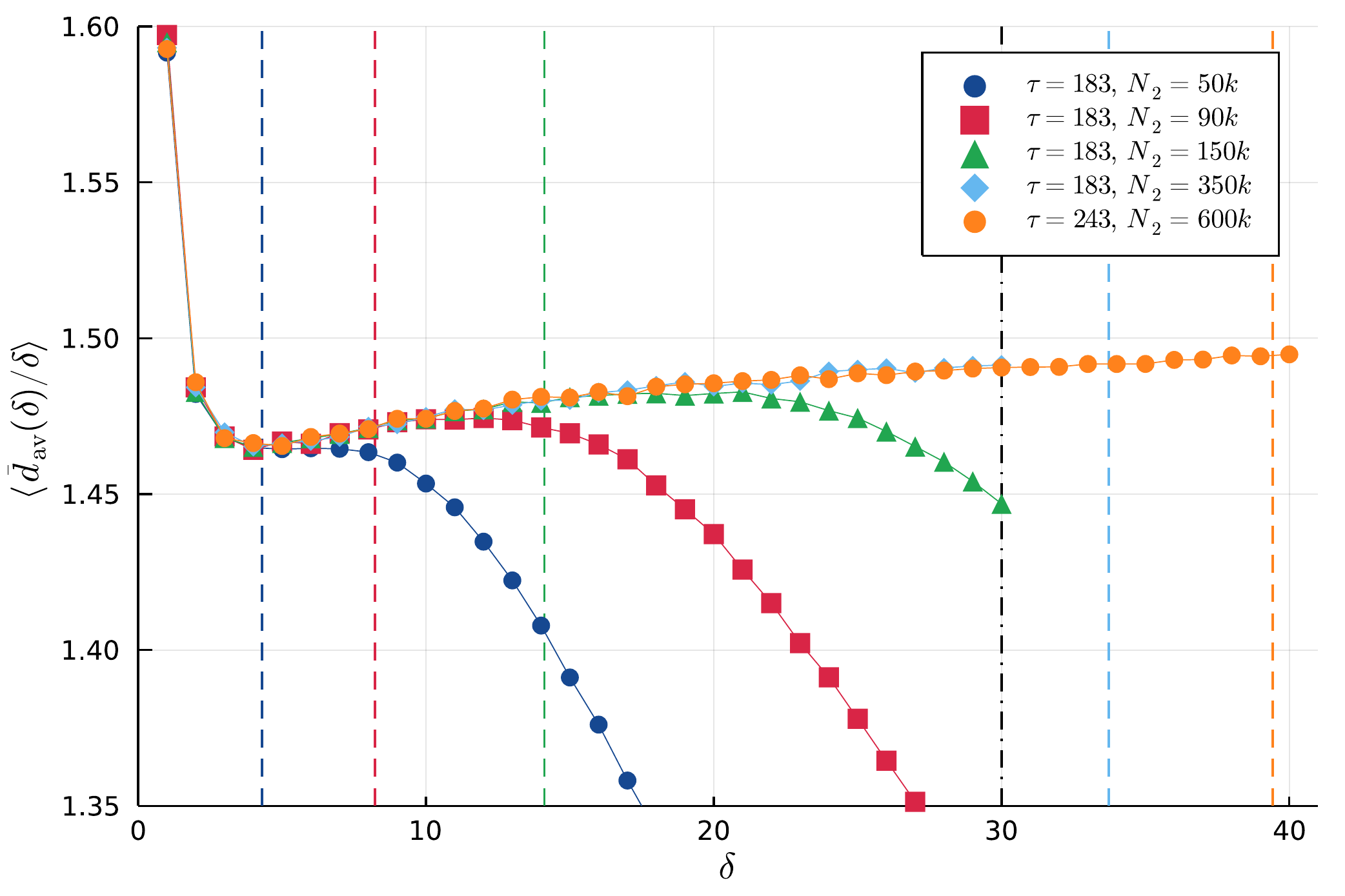}
\caption{Curvature profiles $\langle \bar{d}_{\rm av}(\delta)/\delta \rangle$ measured in 2D CDT quantum gravity on a torus, 
as a function of link distance $\delta$, together with
upper bounds on $\delta$	(vertical dashed lines), indicating the $\delta$-ranges where finite-size effects due to spatial compactness
are negligible \cite{Brunekreef2021}.}
\label{fig:ric2dcdt-delmax}
\end{figure}

The data for the 
curvature profiles are shown in Fig.\ \ref{fig:ric2dcdt-delmax}. As can be seen from the curves for the smaller volumes, their decrease 
for larger $\delta$ seems entirely due to finite-size effects. If one discards the data contaminated by discretization artefacts ($\delta\!\leqslant  \! 5$)
and finite-size effects (to the right of the vertical lines in the figure), they appear to fall on a common curve which increases 
slowly as a function of $\delta$. It is impossible to say whether this curve will eventually asymptote to a flat curve, which would
indicate an average, ``effective" flatness at large coarse-graining scales; there is currently no evidence from available data that this happens. 
Using the dual link distance gives a similar picture. The conclusion at this stage is that the curvature profile of the
2D CDT ``quantum torus", which was dubbed ``quantum-flat'' in \cite{Brunekreef2021}, does not resemble that of any classical, constantly curved 
geometry. As a matter of principle, this is not surprising because of the entirely nonclassical character of quantum gravity in two dimensions.

Interestingly, because of the Lorentzian nature of the CDT geometries and the associated time slicing, one can make use of the directional
nature of the QRC and investigate whether it behaves differently in spacelike and timelike directions. A limited investigation in \cite{Brunekreef2021}
looked at the two cases where the two sphere centres $p$, $p'$ lie in the same spatial slice or, alternatively, have a maximally timelike
separation. The curvature profile for the case of spacelike separation very much resembles that of the directionally averaged data 
of Fig.\ \ref{fig:ric2dcdt-delmax}, however, the curvature profile in the timelike direction has an initial dip (for small $\delta$) that is much less
pronounced than in the other cases and for larger $\delta$ only has a very slight upward slope. Within measuring accuracy it cannot be distinguished 
from a flat curve, indicating a possible anisotropy of the QRC in 2D CDT quantum gravity, which deserves further study.

\begin{figure}[t]
\centerline{\scalebox{0.38}{\rotatebox{0}{\includegraphics{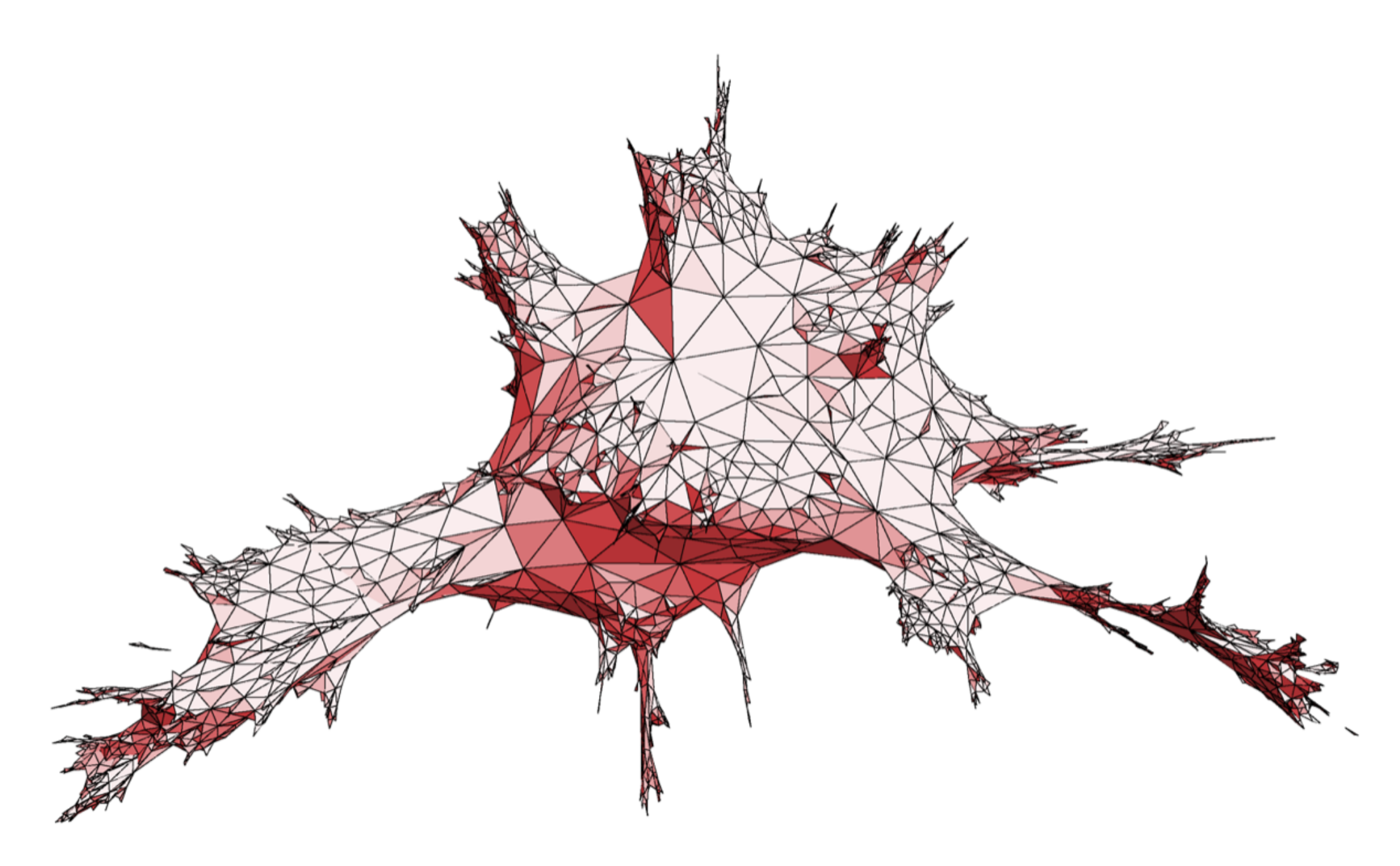}}}}
\caption{Typical configuration in the ensemble of 2D dynamical triangulations of spherical topology. [Figure by courtesy of T.\ Budd]}
\label{fig:DT2Dconfig}
\end{figure}

\subsection{DT quantum gravity in D\,\! =\,\! 2}
\label{sec:dt2}

We turn next to the analysis of the curvature properties of two-dimensional \textit{Euclidean} quantum gravity, as captured by the (continuum limit
of the) nonperturbative DT path integral
\begin{equation}
Z = \sum_{T} \frac{1}{C(T)} \, {\rm e}^{-S_\lambda[T]}, \quad \quad S_\lambda[T] = \lambda\, N_2(T).
\label{path2ddt}
\end{equation}
The difference with the CDT path integral (\ref{path2dcdt}) is the set of two-dimensional triangulations $T$ in the sum, which in (\ref{path2ddt}) 
consists of all
simplicial manifolds of fixed topology that can be assembled from equilateral flat triangles, and is strictly bigger than the corresponding 
configuration space of the Wick-rotated Lorentzian path integral (\ref{path2dcdt}). This difference is significant, since it leads to a different universality 
class of 2D quantum gravity models. More specifically, 
the quantum geo\-metry of 2D Euclidean gravity \cite{David1984,Ambjorn1997,Budd2022}, 
aka Liouville gravity \cite{Mertens2020}, is characterized by a Hausdorff dimension of 4 and a spectral dimension of 2. Typical
path integral configurations are very ``spiky'' and nonclassical, as illustrated by Fig.\ \ref{fig:DT2Dconfig}. 
It seems that before the investigation of the QRC in \cite{Klitgaard2018}, which looked at the standard case with $S^2$-topology, 
no attempt had been made to associate a curvature with this quantum geometry, which in a mathematical context also goes by
the name ``Brownian sphere" \cite{Budd2022}. 

\begin{figure}[t]
\centerline{\scalebox{0.5}{\rotatebox{0}{\includegraphics{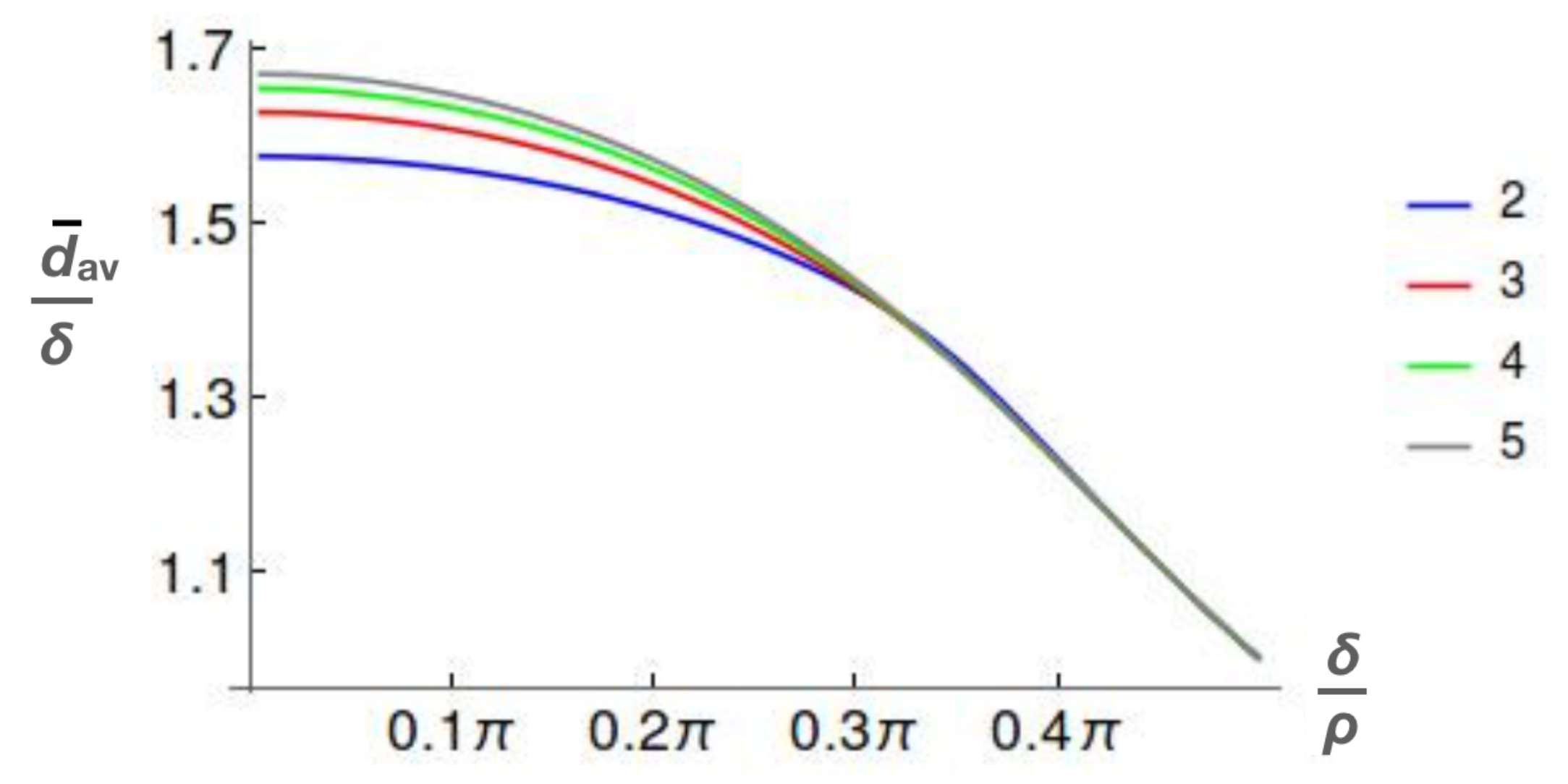}}}}
\caption{Curvature profiles $\bar{d}_{\rm av}(\delta)/\delta$ of continuum spheres in dimensions $D\! =\! 2$, 3, 4 and 5 (bottom to top),
as a function of the normalized distance $\delta/\rho$, where $\rho$ denotes the curvature radius of the sphere \cite{Klitgaard2018}.}
\label{fig:curvspheres}
\end{figure} 

One could of course use the total deficit angle curvature, introduced in Sec.\ \ref{sec:defdt}, but since this is fixed
to a constant by the Gauss-Bonnet theorem, it does not contain any interesting new information. 
By contrast, although the QRC is compatible with the Gauss-Bonnet theorem on triangulations that approximate classical, two-dimensional 
manifolds, for $\delta\!\geq\! 2$ it does in general not satisfy the Gauss-Bonnet theorem.\footnote{Regularity conditions under which the Gauss-Bonnet
theorem holds for $\delta\! =\! 1$ are examined in \cite{Klitgaardthesis}, Sec.\ 3.2.} This is a potential asset from the point of view of the 
nonperturbative quantum theory, since the integrated quantum Ricci scalar is not constrained to behave like that of a two-dimensional
space. It is appropriate in a context where quantum geometry in a continuum limit can scale anomalously, and in particular can have
an effective dimension that is different from the dimension of its elementary building blocks, as is illustrated by 2D Euclidean quantum gravity.

The expectation value $\langle \bar{d}_{\rm av}(\delta)\rangle /\delta$ in 2D DT quantum gravity on a two-sphere was measured in Monte Carlo simulations for
volumes in the range $N_2\!\in\! [20k,240k]$ in \cite{Klitgaard2018}. Since the data for link distances $\delta\! \gtrsim \! 5$ indicate a positive curvature, 
they were fitted to curvature profiles of constantly curved spheres in the continuum. 
In view of the noncanonical scaling properties of the quantum geometry, the dimension of the latter was varied in the range $D\! \in\! [2,5]$ 
(Fig.\ \ref{fig:curvspheres}). From fitting the measurements at a given volume $N_2$ to the curvature profile of a $D$-sphere, 
an effective curvature radius $\rho_{\rm eff}$ was
extracted and used to normalize the distances $\delta$ to $\delta/\rho_{\rm eff}$, for better comparison with the universal curves 
of Fig.\ \ref{fig:curvspheres}. 

\begin{figure}[t]
\centerline{\scalebox{0.55}{\rotatebox{0}{\includegraphics{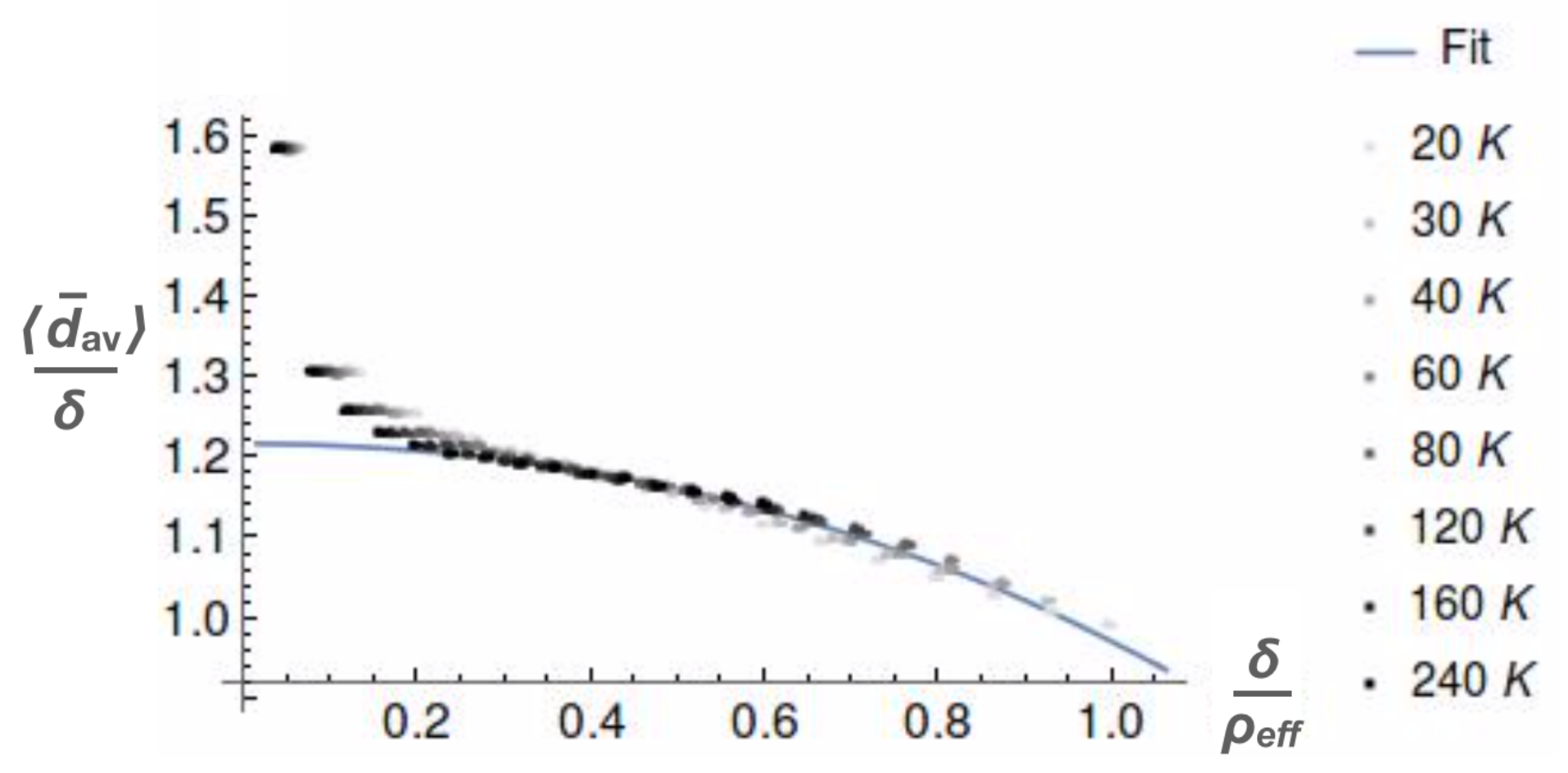}}}}
\caption{Curvature profile $\langle \bar{d}_{\rm av}(\delta)/\delta \rangle$ measured in 2D DT quantum gravity on a sphere as a function of the 
rescaled distance $\delta/\rho_{\mathrm{eff}}$,
for volumes in the range $N_2\!\in\! [20k,240k]$, with a best fit to a 5D continuum sphere \cite{Klitgaard2018}.}
\label{fig:spherefit}
\end{figure} 

Perhaps surprisingly, considering the highly nonclassical nature of the quantum geometry, a joint fit of the
data for different volumes $N_2\!\in\! [20k,240k]$ for a fixed reference dimension $D$ leads to a fairly good overlap with a single continuum 
curve for $\bar{d}_{\rm av}/\delta$ over a whole range of $\delta/\rho_{\rm eff}$-values, indicating the presence of finite-size scaling. 
The quality of the fit improves slightly with the reference dimension and is best for $D\! =\! 5$, the case shown in Fig.\ \ref{fig:spherefit}. 
However, given the similarity of the curves for continuum spheres, Fig.\ \ref{fig:curvspheres}, it is diffcult to discriminate between the cases 
$D\! =\! 4$ and $D\! =\! 5$ on the basis of the sphere fits alone. One can invoke an additional selection criterion by extracting another
effective dimension $\cal D$ from the scaling relation 
\begin{equation}
\rho_{\rm eff} \propto N_2^{1/{\cal D}},
\label{scalerho}
\end{equation}
which determines how the effective curvature radii obtained from the sphere fits at given $D$ depend on the volume $N_2$.
The ansatz (\ref{scalerho}) is motivated by the fact that the curvature radius of a continuum $D$-sphere of two-volume $V_2$ behaves like 
$\rho\!\propto\! V_2^{1/D}$.\
It turns out that, depending on the dimension $D$ and the fitting method used to obtain $\rho_{\rm eff}$, ${\cal D}$ varies in the range $[4.8,6.0]$,
with the closest match between $D$ and $\cal D$ obtained for $D\! =\! 5$ (restricting to integer dimensions). This led to the main conclusion of
\cite{Klitgaard2018}, namely that the curvature profile of 2D Euclidean quantum gravity is best approximated by that of a five-dimensional
continuum sphere. An analysis of the same curvature profile, including on ensembles of triangulations with weaker regularity requirements
than those of simplicial manifolds, which nevertheless are known to lie in the same universality class, are currently underway \cite{new2dcurv}.

\subsection{CDT quantum gravity in D\,\! =\,\! 4}
\label{sec:cdt4}

As already mentioned in the introduction to Sec.\ \ref{sec:qapp}, a main motivation for the introduction of the QRC was the physical case of quantum gravity
in four dimensions, and a closer investigation of the curvature properties of the dynamically generated de Sitter-like quantum geometry found in the 4D CDT
path integral. Its Wick-rotated version was given in eq.\ (\ref{patheu}), and the explicit form of the four-dimensional action is 
\begin{equation}
S[T]\!=\! - \kappa_0 N_0(T) + \Delta (2 N_{41}(T)+N_{32}(T)-6 N_0(T))+\kappa_4 (N_{41}(T)+N_{32}(T)),
\label{action}
\end{equation}
where $N_{41}$ and $N_{32}$ count the numbers of four-simplices of type $(4,1)$ and
$(3,2)$ respectively, with $N_{41}\! +\! N_{32}\! =\! N_4$. These two types correspond to two distinct Minkowskian simplicial building 
blocks before the analytic continuation to
Euclidean signature, and have different numbers of time- and spacelike edges. The bare coupling constants
appearing in the action (\ref{action}) are $\kappa_4$, which is related to the cosmological constant, the inverse Newton constant $\kappa_0$, 
and the asymmetry parameter $\Delta$, which captures the finite relative scaling between the geodesic lengths of
time- and spacelike edges and becomes a relevant coupling in the nonperturbative regime (for more details, see \cite{review1,review2}).
The topology of the triangulations is $S^1\!\times\! S^3$, with compact spatial slices of $S^3$-topology and a compactified time
direction for computational convenience. 

\begin{figure}[t]
\centerline{\scalebox{0.5}{\rotatebox{0}{\includegraphics{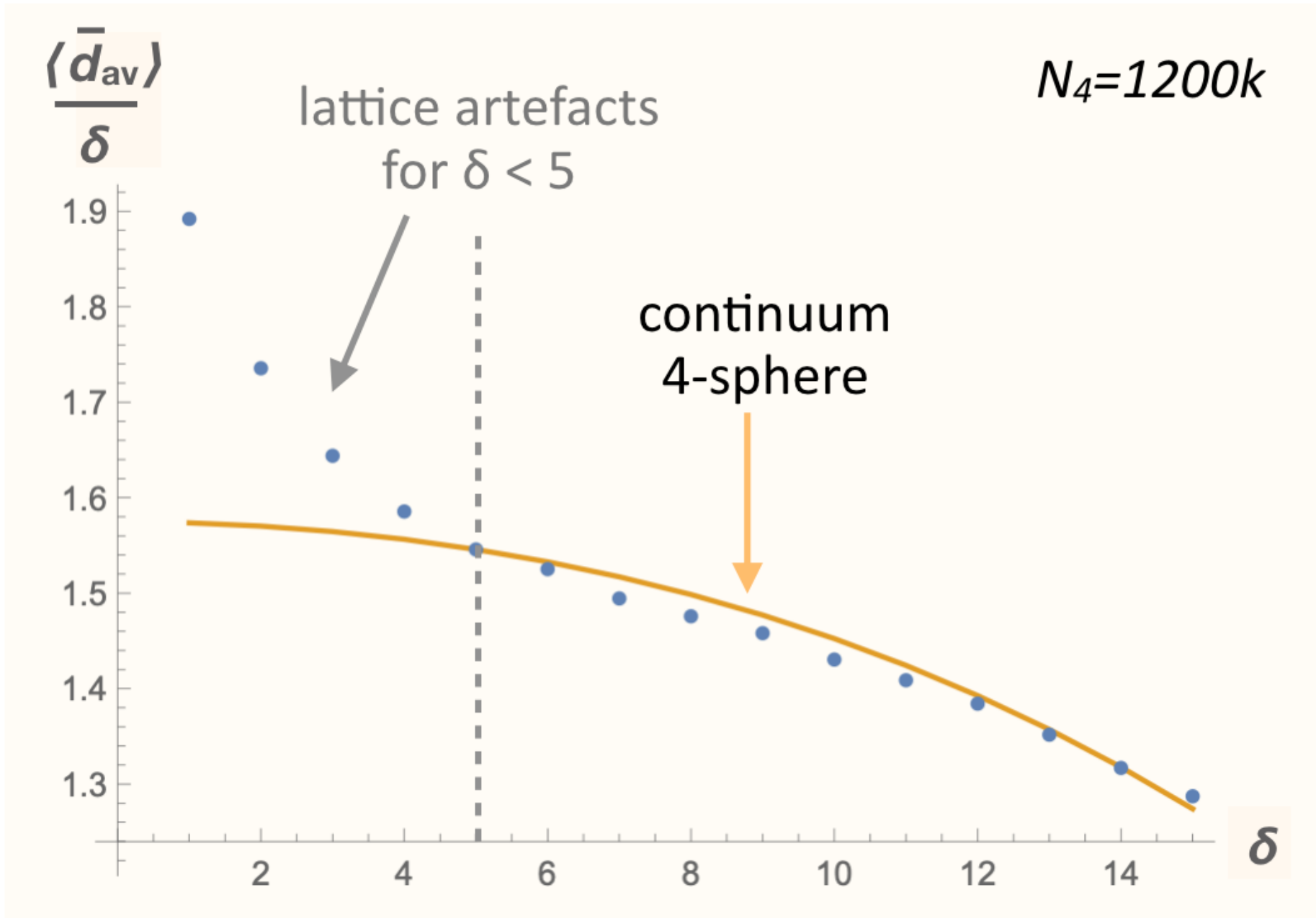}}}}
\caption{Curvature profile $\langle \bar{d}_{\rm av}(\delta)/\delta \rangle$ measured in 4D CDT quantum gravity on $S^1\!\times\! S^3$ in the
de Sitter phase, 
as a function of the dual link distance $\delta$, for volume $N_4\! =\! 1200k$, together with a best fit to the curvature profile of a continuum
four-sphere \cite{Klitgaard2020}.}
\label{fig:curv4d}
\end{figure} 

The Monte Carlo measurements of the curvature profile in \cite{Klitgaard2020} were performed at the point $(\kappa_0,\Delta)\! =\! (2.2,0.6)$ in the 
de Sitter phase of 4D CDT quantum gravity, for universes with time extension $\tau\! =\! 120$, (quasi-)fixed volumes in the range $N_4\! \in\! [150k,1200k]$
and $\delta\! \leqslant \! 15$.
When using the link distance, it was not possible to obtain reliable data for $\langle \bar{d}_{\rm av}(\delta)\rangle /\delta$ because of strong finite-size
effects related to the presence of vertices of large order. The situation improved significantly when using the dual link distance, in the sense that one
could find a nontrivial range of $\delta$-values where the QRC data could be compared meaningfully with
those of continuum spaces, at least for sufficiently large volumes $N_{4}$. The data indicate a positive average quantum Ricci scalar,
suggesting a comparison with curvature profiles for smooth spheres. Unlike in the case of 2D DT quantum gravity described in Sec.\ \ref{sec:dt2},
4D CDT quantum gravity has a nontrivial classical limit, and both Hausdorff and spectral dimension on sufficiently large scales
are compatible with the classically expected value of 4 \cite{Ambjorn2004,Ambjorn2005a,Ambjorn2005}. This suggests a comparison
with the classical curvature profiles of spheres of dimension $D\! =\! 4$. 

Like in the curvature measurements in 2D quantum gravity, the fitting must take into account an (a priori unknown) shift in the measurement data,
corresponding to the non-universal parameter $c_{\rm av}$ in relation (\ref{profile})\footnote{One has the choice between an additive and
a multiplicative shift, which has little effect on the final result. Here, following \cite{Klitgaard2018,Klitgaardthesis}, an additive shift was chosen.},
and data points for $\delta\! <\! 5$ were not considered because of discretization artefacts.
For the smallest investigated four-volumes, the quality of the fits of the 4D data to continuum curves is only moderate.
However, for increasing $N_{4}$ the curve through the measurement data in the range $\delta\!\in\! [5,15]$
becomes gradually more convex, characteristic of the behaviour of a positively curved continuum sphere of radius $\rho$. 
This trend is clearly visible in the CDT data and mirrors the behaviour of the 2D DT system with increasing volume $N_2$,
but the approach to spherical behaviour is slower than in the 2D DT case. 
The quality of the sphere fit at the largest volume $N_{4}\! =\! 1200k$ is decent (Fig.\ \ref{fig:curv4d}), but not as good as the corresponding fit at the 
largest volume $N_2\! =\! 240k$ for the 2D DT measurements. This is not particularly surprising since one would expect
on general grounds that a four-dimensional system shows a slower convergence as
a function of the total volume than a two-dimensional one, even if by some measures two-dimensional quantum gravity 
has an effective dimensionality larger than two. A further consistency check was performed in \cite{Klitgaard2020} by measuring the volumes
of three-dimensional spherical shells and showing that the extracted curvature radius $\rho$ broadly agrees with that obtained from the 
QRC measurements.

To summarize, there is strong evidence from the data collected in the window $\delta\!\in\! [5,15]$
that the expectation value of the average quantum Ricci scalar is compatible with that of a classical four-sphere. 
Given that the physical length scale at which the curvature is probed is not more than about 10 Planck lengths \cite{Klitgaard2020},
this is quite a remarkable result, and a further piece of evidence for the de Sitter nature of the dynamically generated quantum 
geometry in 4D CDT quantum gravity. In addition, analogous to what was described for 2D CDT in Sec.\ \ref{sec:cdt2}, the directional
character of the QRC was employed to compare the behaviour of the curvature profile in a maximally timelike and maximally spacelike
direction, adapted to the formulation on the dual lattice. The result of this preliminary study, unlike that in two dimensions, found no
significant difference between the time- and spacelike measurements, other than a constant shift, corresponding to different constants
$c_{\rm av}$ associated with the anisotropy of the underlying lattice structure, which was already noted for flat lattices in Sec.\ \ref{sec:constr}.  
It suggests a restoration of local ``rotational" symmetry, which is present on a classical de Sitter space.

\section{Summary and outlook}
\label{sec:future}

As we have seen in Secs.\ \ref{sec:qrc} and \ref{sec:qapp}, the quantum Ricci curvature is a well-defined notion of Ricci curvature,
which has been devised for application in nonperturbative quantum gravity, where typical spacetime configurations contributing to the
path integral are neither classical nor smooth. The QRC $K_q$ at scale $\delta$ of eq.\ (\ref{ricdefine}) is defined operationally in terms of distance and volume
measurements, which is a typical feature of (ingredients of) observables in a nonperturbative regime. It provides a benchmark for
the type of curvature information one is able to access in such a regime, which is dictated not only by the highly nonclassical
character of the geometry, but also by the requirement of diffeomorphism-invariance and background-independence, typically implying
some form of integration or averaging. The message here is that natural, well-defined quantum operators are not of the form of individual quantized components 
$\hat{R}^\kappa_{\lambda\mu\nu}(x)$ of the classical Riemann tensor, but are nonlocal, composite operators like the QRC.

The QRC is a ``tensorial" quantity in the sense that it encodes information beyond the Ricci scalar, depending
on $Ric(v,v)$ and its covariant derivatives, as is illustrated by the expansions (\ref{rnc}) for the average sphere distance on Riemannian spaces
for infinitesimal $\delta$. This direction-dependence has so far only been used to a limited extent in quantum applications\footnote{In a classical
context, it has been used to investigate the anisotropic curvature properties of a two-dimensional ellipsoid \cite{Klitgaardthesis,newellipsoid}.}, 
namely in CDT models
to get a first idea of the dependence of the QRC on time- vs.\ spacelike directions, as men\-tioned in Secs.\ \ref{sec:cdt2} and \ref{sec:cdt4} above. 
This feature will become more prominent in upcoming
studies of the curvature properties of coupled gravity-matter systems, including point particles. 

The simplest
diffeomorphism-invariant observable constructed from the QRC is the curvature profile (\ref{profile}), which only depends on the Ricci scalar, 
at least at lowest nontrivial order
in $\delta$.\footnote{Since the standard prescription for the QRC for $\epsilon\! =\! \delta$ (cf.\  Sec.\ \ref{sec:constr}) is not rotationally symmetric, it still
has a directional character; however, choosing $\delta\! =\! 0$ is associated with full rotational symmetry.} Importantly, although the curvature profile is an
integrated, global (albeit scale-dependent) observable, its application in full 4D CDT quantum gravity has already led to two major new results.
Firstly, it implies a finite, renormalized average curvature on higher-dimensional systems of dynamical triangulations, 
unlike the deficit angle curvature of Sec.\ \ref{sec:defdt}. 
Secondly, it has produced additional evidence that the 
classical limit of the quantum geo\-metry found in this candidate theory of quantum gravity is indeed a de Sitter space.

Overall, the advent of a well-defined quasi-local notion of curvature in nonperturbative quantum gravity implies a major advance, which is likely to be 
transferable to approaches different from CDT, if they have suitable computational means to implement the QRC. Its
computational implementation and measurement, especially in higher dimensions, is fairly complex, and any further numerical optimization will be welcome.  
The QRC offers a plethora of possibilities for constructing new observables, since the only (quasi-)local physical quantities available to date
were the Hausdorff and spectral dimensions. Quantities involving local vertex orders (numbers of $d$-simplices meeting at a vertex), closely
related to deficit angles, have also been used on occasion, but their relation to any physical, renormalized notion of curvature is tenuous. 

Examples of new observables where the local QRC is expected to play a prominent role are homogeneity measures of the type introduced
in \cite{Loll2022} and two-point functions \cite{new2point}, earlier studied in the context of Euclidean dynamical 
triangulations \cite{deBakker1995,Ambjorn1998a}. 
Determining the degree of homogeneity and isotropy of the de Sitter-like quantum spacetime will be an essential step in trying to relate it
to continuum descriptions of the very early universe, where these symmetries are usually \textit{assumed} to be present. The behaviour of two-point 
functions captures essential spacetime properties of any quantum field theory, while two-point functions of spatial slices
of a cosmological background are closely related to the power spectrum \cite{Ellis2012}.
In other words, we are interested in both spacetime curvature and in the spatial curvature of slices of constant 
time\footnote{The curvature profile of two-dimensional
spatial slices in three-dimensional CDT quantum gravity was investigated in \cite{Brunekreef2022,Brunekreefthesis}.} or of other subspaces.
Many interesting applications remain to be explored, given the many invariant aspects of classical General Relativity 
that are expressed in terms of curvature and the fact that we have only just begun to understand the
true quantum properties of curvature in a Planckian regime.

\vspace{0.2cm}

\subsection*{Acknowledgments} 

\noindent I am grateful to numerous individuals for discussion and collaboration on various aspects of quantum curvature,
in particular, N.\ Klitgaard, J.\ Brunekreef, G.\ Clemente, J.\ Ambj\o rn, T.\ Niestadt, A.\ Silva  and J.\ van der Duin.
This research was supported in part by Perimeter Institute for Theoretical Physics. Research at Perimeter Institute is supported by the 
Government of Canada through the Department of Innovation, Science and Economic Development and by the Province of Ontario through
the Ministry of Colleges and Universities.

\end{document}